\documentclass[aps,prd,twocolumn,floatfix,preprintnumbers,nofootinbib,superscriptaddress]{revtex4}
\usepackage{amsmath,amsthm,amssymb,amsfonts}
\usepackage{graphicx}
\usepackage{textcomp}
\usepackage{bm}
\usepackage{color}
\usepackage{multirow}
\usepackage{dutchcal}
\usepackage{makecell}

\usepackage[outdir=./]{epstopdf}

\begin{document}

\title{The low-lying hidden- and double-charm tetraquark states in a constituent quark model with Instanton-induced Interaction}

\author{Jin-Bao Wang}
\affiliation{School of Physical Science and Technology, Southwest
University, Chongqing 400715, China}

\author{Gang Li}~\email{gli@qfnu.edu.cn}
\affiliation{College of Physics and Engineering, Qufu Normal
University, Qufu 273165, China}

\author{Chun-Sheng An}~\email{ancs@swu.edu.cn}
\affiliation{School of Physical Science and Technology, Southwest
University, Chongqing 400715, China}

\author{Cheng-Rong Deng}
\affiliation{School of Physical Science and Technology, Southwest
University, Chongqing 400715, China}

\author{Ju-Jun Xie}
\affiliation{Institute of Modern Physics, Chinese Academy of
	Sciences, Lanzhou 730000, China} \affiliation{School of Nuclear
	Science and Technology, University of Chinese Academy of Sciences,
	Beijing 101408, China} \affiliation{School of Physics and
	Microelectronics, Zhengzhou University, Zhengzhou, Henan 450001,
	China} \affiliation{Lanzhou Center for Theoretical Physics, Key
	Laboratory of Theoretical Physics of Gansu Province, Lanzhou
	University, Lanzhou 730000, China}

\date{\today}

\begin{abstract}
Spectrum of the low-lying hidden- and double-charm tetraquark states are investigated in a nonrelativistic quark potential model, where the Instanton-induced interaction is taken as the residual spin-dependent hyperfine interaction between quarks. The model parameters are fixed by fitting the spectrum of the ground hadron states. Our numerical results show that masses of several presently studied tetraquark states are close to those of the experimentally observed candidates of exotic meson, which indicates that the corresponding compact tetraquark components may take considerable probabilities in those observed exotic states.
	
\end{abstract}

\maketitle

\section{Introduction}

More and more exotic hadron candidates, which exhibit different properties from mesons and baryons in the traditional quark model, have been observed experimentally~\cite{PDG} since the discovery of $X(3872)$~\cite{Belle 2003}.
The typical mesonic exotic states, such as the charmonium-like state $Z_{c}(3900)^{\pm}$ observed in the $\pi^{\pm}J/\psi$ invariant mass spectra of the process $e^+e^-\rightarrow\pi^+\pi^-J/\psi$~\cite{BESIII 2013} and the $Z_{cs}(3985)$ state observed in the $D_s^-D^{*0}$ and $D_s^{*-}D^0$ final states of the $e^+e^-\rightarrow K^+D_s^-D^{*0}(D_s^{*-}D^0)$ reaction~\cite{BESIII 2021}, may contain at least four $c\bar{c}q\bar{q}$ quarks. Theoretically, the exotic meson states can be explained as compact tetraquark states in quark potential model~\cite{Yang:2009zzp,Deng:2016rus,Richard:2017una,Luo:2017eub,JinX 2021,YangG 2021,Deng 2021}, QCD sum rule~\cite{Chen:2013pya,Chen:2015ata,Wang:2017dtg,Wang:2019got}, and diquark-diquark picture~\cite{D.Ebert 2007,Esposito:2013fma,Shi:2021jyr}. While, they could be also described as the molecular states in effective field theory~\cite{Liu:2009qhy,WangP 2013,HeJ 2013,Aceti:2014kja,ChenR 2021,Feijoo:2021ppq}. See also Refs.~\cite{Chen:2016qju,Lebed:2016hpi,Guo:2017jvc,Liu:2019zoy,Brambilla:2019esw} for recent reviews.

Very recently, an exotic state $T_{cc}^{+}$ was reported in the $D^0D^0\pi^+$ mass spectrum by LHCb collaboration~\cite{LHCbTcc1,LHCbTcc2}, which has a mass very close to the $D^0D^{*+}$ threshold and its width is extremely narrow. In fact, taking into account the heavy quark symmetry between $\Xi_{cc}$ and $T_{cc}$, the existence of tetraquark states which contain two heavy quarks has been predicted theoretically in Refs.~\cite{Zhu 2013,Kar.M 2017}, after the discovery of double-charm baryon $\Xi_{cc}^{++}$~\cite{Xicc}. Especially, the binding energy of $DD^*$ molecule predicted by one boson exchange model is in agreement with the experimental value. Within the $DD^*$ molecule nature, the strong and radiative decay widths of $T^+_{cc}$ are studied in Ref.~\cite{Feijoo:2021ppq,Ling:2021bir}. The successful prediction and the discovery of $T_{cc}^{+}$ are enormously helpful to our understanding of the hadron structure.

The constituent quark model~(CQM), in which different versions of the explicit hyperfine interactions between quarks have been proposed, is one of the most successful phenomenological method to study the hadron structure. For instance, in the Godfrey-Isgur model~\cite{Isgur}, all the effects of one gluon exchange interactions (OGE) are included, and it gives a good description for meson spectra from $\pi$ to $\Upsilon$ meson. The chiral constituent quark model~\cite{Glozman:1995fu,Glozman:1995xy,Vijande:2004he,Deng:2014gqa,Yang:2015bmv,Deng:2017xlb} is another widely used version to investigate hadron spectrum, which usually contains one boson exchange~(OBE) interactions, characterizing chiral symmetry spontaneously broken of QCD at low energy.
Tentatively, different versions of CQM have been compared in the estimations on spectrum of baryon excitations
in a multiquark picture recently~\cite{Yuan:2012wz,Yuan:2012zs,An:2013zoa,An:2014lga,An:2017lwg,Wang:2021rjk}.

In addition to the widely employed OGE- and OBE-CQM, the instanton-induced interaction between quarks has also been
used in Refs.~\cite{Yuan:2012wz,Yuan:2012zs,An:2013zoa} to investigate the spectrum of baryon excitations as pentaquark sates. This kind of interaction was firstly introduced by 't Hooft to solve the $U(1)$-problem~\cite{tHooft}, and the instanton vacuum of QCD could provide a mechanism of spontaneous chiral symmetry breaking of QCD~\cite{Diakonov:1983hh,Diakonov:1985eg}. Phenomenologically, it has been shown that one can successfully reproduce the spectrum of light hadrons~\cite{Shuryak:1988bf,Blask:1990ez,Brau:1998sxe,Semay:2001th} and the single charm baryons~\cite{Migura:2006ep} employing the instanton-induced interaction. Furthermore, the instanton-induced interaction has also been employed to study the tetraquark states~\cite{Metch,tHooft:2008rus}. Consequently, here we use it to investigate the spectra of low-lying hidden- and double-charm tetraquark states.

The present manuscript is organized as follows. In Sec.~\ref{Theory}, we present the formalism of the CQM with instanton-induced interaction between quarks, including effective Hamiltonian, wave functions for tetraquark configurations and model parameters. The masses of hidden- and double-charm tetraquark states in our model and discussions are given in Sec.~\ref{ResultsAndDiscussions}. Finally, Sec.~\ref{Summary} is a brief summary of present manuscript.

\section{Framework}
\label{Theory}

\subsection{Effective Hamiltonian}

In present work, a non-relativistic quark potential model is employed for calculations on the spectra of hidden- and double-charm tetraquark states. The dynamics of the model are governed by an effective Hamiltonian as follow:
\begin{equation}
H_{eff.}=\sum_{i=1}^{4}\left(m_i+T_i\right)-T_{C.M.} + V_{Conf.}+V_{Ins.}\,,\label{H}
\end{equation}
where $m_i$ and $T_i$ are the constituent mass and kinetic energy of $i$-th quark, $T_{C.M.}$ denotes center of mass kinetic energy. $V_{Conf.}$ stands for the quark confinement potential, which is taken to be the widely accepted Cornell potential in present work~\cite{Cornell}:
\begin{equation}
V_{Conf.}=\sum_{i<j}-\frac{3}{16}\,\left(\vec{\lambda}^c_i\cdot\vec{\lambda}^c_j\right)\,\left(b\,r_{ij}
-\frac{4}{3}\frac{\alpha_{ij}}{r_{ij}}+C_0\right)\,,
\end{equation}
where $\vec{\lambda}^c_{i(j)}$ is Gell-Mann matrix in $SU(3)$ color space acting on the $i(j)$-th quark, $b$, $\alpha_{ij}$ and $C_0$  are strength of quark confinement, QCD effective coupling constant between two quarks and zero point energy, respectively.

For the residual spin-dependent interaction $V_{Ins.}$ in Eq.~\eqref{H}, we employ a phenomenologically extended version of 't Hooft's instanton-induced interaction~\cite{Migura:2006ep}:
\begin{equation}
V_{Ins.}=V^{qq}_{Ins.}+V^{q\bar{q}}_{Ins.}\,,
\end{equation}
with
\begin{eqnarray}
V^{qq}_{Ins.}&=&\sum_{i<j}-\hat{g}^{qq}_{ij}\Big(P_{ij}^{S=1}P_{ij}^{C,\bf{6}}+2P_{ij}^{S=0}
P_{ij}^{C,\bf{\bar{3}}}\Big)\delta^3(\vec{r}_{ij})\,,\nonumber\\
V^{q\bar{q}}_{Ins.}&=&\sum_{i<j}\hat{g}^{q\bar{q}}_{ij}\Bigg[\frac{3}{2}P_{ij}^{S=1}P_{ij}^{C,\bf{8}}+P_{ij}^{S=0}
\Big(\frac{1}{2}P_{ij}^{C,\bf{8}}\nonumber\\
&&+8P_{ij}^{C,\bf{1}}\Big)\Bigg]\delta^3(\vec{r}_{ij})\,,
\end{eqnarray}
where $V^{qq}_{Ins.}$ is for the interaction between a quark-quark or antiquark-antiquark pair,
and $V^{q\bar{q}}_{Ins.}$ is for that between a quark-antiquark pair.
$\hat{g}^{qq}_{ij}$ and $\hat{g}^{q\bar{q}}_{ij}$ are flavor-dependent coupling strength operators, whose explicit matrix elements are shown in Table~\ref{Tgqq} and Table~\ref{Tgqqbar}. $P_{ij}^{S=0}$ and $P_{ij}^{S=1}$ are spin projector operators onto spin-singlet and spin-triplet states, respectively;
$P_{ij}^{C,\bf{\bar{3}}}$, $P_{ij}^{C,\bf{6}}$, $P_{ij}^{C,\bf{1}}$ and $P_{ij}^{C,\bf{8}}$ are color projector operators onto color anti-triplet $\bf{\bar{3}}_c$, color sextet $\bf{6}_c$, color singlet $\bf{1}_c$ and color octet $\bf{8}_c$, respectively.

One should note that instanton-induced interaction is a pure contact interaction, and $\delta^3\left(\vec{r}_{ij}\right)$ will lead to an unbound Hamiltonian if we don't treat it perturbatively. Here we regularize it as in Refs.~\cite{Isgur,Vijande:2004he,Metch},
\begin{equation}
\delta^3\left(\vec{r}_{ij}\right)\,\rightarrow\,\left(\frac{\sigma}{\sqrt{\pi}}\right)^3\mathrm{exp}\left(-\sigma^2\,r_{ij}^2\right)\,.
\end{equation}
The regularization parameter $\sigma$ should correlate to the finite size of constituent quarks and therefore flavor dependent~\cite{Vijande:2004he}, but here we just make $\sigma$ to be a unified parameter for different quark flavors and absorb the flavor dependence into $\hat{g}^{qq}_{ij}$ and $\hat{g}^{q\bar{q}}_{ij}$.

\begin{table*}[h]
	\caption{Matrix elements of flavor-dependent coupling strength operators $\hat{g}^{qq}_{ij}$. $g_{f_1f_2}$ is flavor-dependent model parameter, $n=u,d$ stands for light flavor.}
	\centering
	\renewcommand
	\tabcolsep{6pt}
	\renewcommand{\arraystretch}{1.3}
	\label{Tgqq}
	\begin{tabular}{c|cc|cccc|cccc|cc|cccc}
		\hline\hline
		&$ud     $&$     du$&$     us$&$ds     $&$sd     $&$su     $&$uc$&$dc$&$cd$&$cu$&$sc$&$cs$&$uu$&$dd$&$ss$&$cc$\\
		\hline
		$ud$&$ g_{nn}$&$-g_{nn}$&\multicolumn{4}{c|}{\multirow{2}{*}{0}}&\multicolumn{4}{c|}{\multirow{2}{*}{0}}&\multicolumn{2}{c|}{\multirow{2}{*}{0}}&\multicolumn{4}{c}{\multirow{2}{*}{0}}\\
		
		$du$&$-g_{nn}$&$ g_{nn}$&    &    &   &    &     &    &   &    &    &    &    &    &     &     \\
		\hline
		$us$&\multicolumn{2}{c|}{\multirow{4}{*}{0}}&$ g_{ns}$&$  0     $&$ 0      $&$-g_{ns}$&\multicolumn{4}{c|}{\multirow{4}{*}{0}}&\multicolumn{2}{c|}{\multirow{4}{*}{0}}&\multicolumn{4}{c}{\multirow{4}{*}{0}}\\
		
		$ds$&         &         &$  0     $&$ g_{ns}$&$-g_{ns}$&$   0    $&     &    &    &    &     &    &&&& \\
		
		$sd$&         &         &$   0    $&$-g_{ns}$&$ g_{ns}$&$   0    $&     &    &    &    &     &    &&&& \\
		
		$su$&         &         &$-g_{ns}$&$    0   $&$     0  $&$ g_{ns}$&     &    &    &    &     &    &&&& \\
		\hline
		$uc$&\multicolumn{2}{c|}{\multirow{4}{*}{0}}&\multicolumn{4}{c|}{\multirow{4}{*}{0}}&$g_{nc}$& 0 & 0 &$-g_{nc}$&\multicolumn{2}{c|}{\multirow{4}{*}{0}}&\multicolumn{4}{c}{\multirow{4}{*}{0}} \\
		
		$dc$&      &           &        &            &      &          & 0 &$g_{nc}$&$-g_{nc}$& 0 &   & &   &  &  & \\
		
		$cd$&      &           &        &            &      &          & 0 &$-g_{nc}$&$g_{nc}$& 0 &   & &   &  &  & \\
		
		$cu$&      &           &        &            &      &          &$-g_{nc}$& 0 & 0 &$g_{nc}$&   & &   &  &  & \\
		\hline
		$sc$&\multicolumn{2}{c|}{\multirow{2}{*}{0}}&\multicolumn{4}{c|}{\multirow{2}{*}{0}}&\multicolumn{4}{c|}{\multirow{2}{*}{0}}   &$g_{sc}$&$-g_{sc}$&\multicolumn{4}{c}{\multirow{2}{*}{0}}\\
		
		$cs$&     &            &      &            &         &       &          &   &  &        &$-g_{sc}$&$g_{sc}$&&&&\\
		\hline
		$uu$&\multicolumn{2}{c|}{\multirow{4}{*}{0}}&\multicolumn{4}{c|}{\multirow{4}{*}{0}}&\multicolumn{4}{c|}{\multirow{4}{*}{0}}      &\multicolumn{2}{c|}{\multirow{4}{*}{0}}&\multicolumn{4}{c}{\multirow{4}{*}{0}}\\
		
		$dd$&     &            &      &            &         &       &          &   &  &        & & &&&&\\
		
		$ss$&     &            &      &            &         &       &          &   &  &        & & &&&&\\
		
		$cc$&     &            &      &            &         &       &          &   &  &        & & &&&&\\
		\hline\hline		
	\end{tabular}	
\end{table*}

\begin{table*}[h]
	\caption{Matrix elements of flavor-dependent coupling strength operator $\hat{g}^{q\bar{q}}_{ij}$. $g_{f_1f_2}$ is flavor-dependent model parameter, $n=u,d$ stands for light flavor.}
	\centering
	\renewcommand
	\tabcolsep{5.8pt}
	\renewcommand{\arraystretch}{1.3}
	\label{Tgqqbar}
	\begin{tabular}{c|cc|cccc|cccc|cc|cccc}
		\hline\hline
		&$u\bar{d}$&$d\bar{u}$&$u\bar{s}$&$d\bar{s}$&$s\bar{d}$&$s\bar{u}$&$u\bar{c}$&$d\bar{c}$&$c\bar{d}$&$c\bar{u}$&$s\bar{c}$&$c\bar{s}$&$u\bar{u}$&$d\bar{d}$&$s\bar{s}$&$c\bar{c}$\\
		\hline
		$u\bar{d}$&$-g_{nn}$&$0$&\multicolumn{4}{c|}{\multirow{2}{*}{0}}&\multicolumn{4}{c|}{\multirow{2}{*}{0}}&\multicolumn{2}{c|}{\multirow{2}{*}{0}}&\multicolumn{4}{c}{\multirow{2}{*}{0}}\\
		
		$d\bar{u}$&$0$&$-g_{nn}$&&&&&&&&&&&&&&\\
		\hline
		$u\bar{s}$&\multicolumn{2}{c|}{\multirow{4}{*}{0}}&$-g_{ns}$&0&0&0&\multicolumn{4}{c|}{\multirow{4}{*}{0}}&\multicolumn{2}{c|}{\multirow{4}{*}{0}}&\multicolumn{4}{c}{\multirow{4}{*}{0}}\\
		
		$d\bar{s}$& &  &0&$-g_{ns}$&0&0&&&&&&&&&&\\
		
		$s\bar{d}$& &  &0&$0$&$-g_{ns}$&0&&&&&&&&&&\\
		
		$s\bar{u}$&&   &0&$0$&$0$&$-g_{ns}$&&&&&&&&&&\\
		\hline
		$u\bar{c}$&\multicolumn{2}{c|}{\multirow{4}{*}{0}}&\multicolumn{4}{c|}{\multirow{4}{*}{0}}&$-g_{nc}$&0&0&0&\multicolumn{2}{c|}{\multirow{4}{*}{0}}&\multicolumn{4}{c}{\multirow{4}{*}{0}}\\
		
		$d\bar{c}$& &   &  &  &  &   &$0$&$-g_{nc}$&0&0&&&&&&\\
		
		$c\bar{d}$& &   &  &  &  &   &$0$&$0$&$-g_{nc}$&0&&&&&&\\
		
		$c\bar{u}$& &   &  &  &  &   &$0$&$0$&$0$&$-g_{nc}$&&&&&&\\
		\hline
		$s\bar{c}$&\multicolumn{2}{c|}{\multirow{2}{*}{0}}&\multicolumn{4}{c|}{\multirow{2}{*}{0}}&\multicolumn{4}{c|}{\multirow{2}{*}{0}}&$-g_{sc}$&0&\multicolumn{4}{c}{\multirow{2}{*}{0}}\\
		
		$c\bar{s}$& &   &  &  &  &   &   &   &   &   &$0$&$-g_{sc}$&&&&\\
		\hline
		$u\bar{u}$&\multicolumn{2}{c|}{\multirow{4}{*}{0}}&\multicolumn{4}{c|}{\multirow{4}{*}{0}}&\multicolumn{4}{c|}{\multirow{4}{*}{0}}&\multicolumn{2}{c|}{\multirow{4}{*}{0}}&0&$g_{nn}$&$g_{ns}$&$g_{nc}$\\
		
		$d\bar{d}$& &   &  &  &  &   &   &   &   &   &    &    &$g_{nn}$&$0$&$g_{ns}$&$g_{nc}$\\
		
		$s\bar{s}$& &   &  &  &  &   &   &   &   &   &    &    &$g_{ns}$&$g_{ns}$&$0$&$g_{sc}$\\
		
		$c\bar{c}$& &   &  &  &  &   &   &   &   &   &    &    &$g_{nc}$&$g_{nc}$&$g_{sc}$&$0$\\
		\hline\hline
	\end{tabular}
\end{table*}	

\subsection{Configurations of four quark systems}

Considering the Pauli principle for two-quark (-antiquark) subsystems, there are eight possible
flavor-spin-color symmetry configurations for $qq\bar{q}\bar{q}$ systems, as follows:
\begin{flalign}
\hspace{1cm}|1\rangle&=\{qq\}_{{\bf{6}}_c}\{\bar{q}\bar{q}\}_{\bar{\bf{6}}_c}\,, & |2\rangle&=[qq]_{\bar{\bf{3}}_c}[\bar{q}\bar{q}]_{{\bf{3}}_c}\,,\hspace{1cm}\notag\\
|3\rangle&=\{qq\}^{*}_{\bar{\bf{3}}_c}[\bar{q}\bar{q}]_{{\bf{3}}_c}\,,& |4\rangle&=[qq]_{\bar{\bf{3}}_c}\{\bar{q}\bar{q}\}^{*}_{{\bf{3}}_c}\,,\notag\\
|5\rangle&=[qq]^{*}_{{\bf{6}}_c}\{\bar{q}\bar{q}\}_{\bar{\bf{6}}_c}\,,& |6\rangle&=\{qq\}_{{\bf{6}}_c}[\bar{q}\bar{q}]^{*}_{\bar{\bf{6}}_c}\,,\notag\\
|7\rangle&=\{qq\}^{*}_{\bar{\bf{3}}_c}\{\bar{q}\bar{q}\}^{*}_{{\bf{3}}_c}\,,& |8\rangle&=[qq]^{*}_{{\bf{6}}_c}[\bar{q}\bar{q}]^{*}_{\bar{\bf{6}}_c}\,,\label{SymConfig}
\end{flalign}
where $\{\dots\}$ denotes a permutation symmetric flavor wave function and $[\dots]$ denotes a permutation antisymmetric flavor wave function of two-quark (-antiquark) subsystem.
The superscript $*$ means diquark or antidiquark forms a spin-triplet, while the configurations without superscript
are spin-singlet.
The blackened number in the subscript denotes color wave function of two-quark (-antiquark) subsystem. As we can see, there are two possible color neutral structures of tetraquark systems, namely, $\bf{6}_c\otimes\bf{\bar{6}}_c$ and $\bf{\bar{3}}_c\otimes\bf{3}_c$.
Note that here we only consider the $S$-wave tetraquark systems, so the spatial wave functions are always completely
permutation symmetric.

Explicit flavor wave functions of the studied hidden- and double-charm tetraquark systems are presented in the following
subsections, respectively.

\subsubsection{Hidden-Charm Systems}
\label{fhc}

There are nine flavor configurations for $q\bar{q}c\bar{c}$ with $q$ the light quarks $u$, $d$, and $s$, which are:
$uc\bar{u}\bar{c}$, $uc\bar{d}\bar{c}$, $uc\bar{s}\bar{c}$, $dc\bar{d}\bar{c}$, $dc\bar{s}\bar{c}$, $sc\bar{s}\bar{c}$, $dc\bar{u}\bar{c}$, $sc\bar{u}\bar{c}$, and $sc\bar{d}\bar{c}$. Compositions of these configurations could lead to the tetraquark states with quantum numbers
isospin $I$ and strangeness $S$, as below:
\begin{itemize}
	\item $I=0,\,S=0.$
	\begin{align}
	X_{n\bar{n}}&=\frac{1}{\sqrt{2}}\left(uc\bar{u}\bar{c}+dc\bar{d}\bar{c}\right),\label{X1f}\\
	X_{s\bar{s}}&=sc\bar{s}\bar{c}.\label{X2f}
	\end{align}
	\item $I=\frac{1}{2},\,S=\pm1.$
	\begin{align}
	&Z_{cs}^{+}=uc\bar{s}\bar{c},&&\bar{Z}_{cs}^{0}=-sc\bar{d}\bar{c},\label{fZcs+}\\
	&Z_{cs}^{0}=dc\bar{s}\bar{c}.&&Z_{cs}^{-}=sc\bar{u}\bar{c}.\label{fZcs0}
	\end{align}
	\item $I=1,\,S=0.$
	\begin{align}
	Z_{c}^{+}&=-uc\bar{d}\bar{c},\label{Zc+}\\
	Z_{c}^{0}&=\frac{1}{\sqrt{2}}\left(uc\bar{u}\bar{c}-dc\bar{d}\bar{c}\right),\label{Zc0}\\
	Z_{c}^{-}&=dc\bar{u}\bar{c}.\label{Zc-}
	\end{align}
\end{itemize}
$X$ and $Z_c^{0}$ are pure neutral systems, thus they could have $C$-parity. Note that $\bar{Z}_{cs}^{0}$ and $Z_{cs}^{-}$ are the conjugate states of $Z_{cs}^{0}$ and $Z_{cs}^{+}$, respectively. Thus, we only consider the latter set in following, since a state should share the same energy with its conjugate state in present model.

Accordingly, the symmetry configurations of hidden-charm systems in Eq.~(\ref{SymConfig}) could be classified by quantum numbers $J^{P(C)}$, which are listed in Table~\ref{HCSym}.

\begin{table}[htbp]
\caption{The symmetry configurations of hidden-charm systems.}\label{HCSym}
\renewcommand
\tabcolsep{0.35cm}
\renewcommand{\arraystretch}{2}
\begin{tabular}{cccc}
\hline\hline
\multicolumn{2}{c}{For $Z_{cs}$ states}&\multicolumn{2}{c}{For $X$ and $Z_{c}$ states} \\

$J^{P}$&Configuration&$J^{PC}$&Configuration\\\hline

$0^+ $&  $|1\rangle$ &$0^{++} $&  $|1\rangle$ \\

      &  $|2\rangle$ &      &  $|2\rangle$ \\

      &  $|7\rangle$ &      &  $|7\rangle$ \\

      &  $|8\rangle$ &      &  $|8\rangle$ \\\hline

$1^+$ &  $|3\rangle$ &$1^{++}$ &  $|3'\rangle=\frac{1}{\sqrt{2}}\left(|3\rangle+|4\rangle\right)$ \\

      &  $|4\rangle$ &      &  $|5'\rangle=\frac{1}{\sqrt{2}}\left(|5\rangle+|6\rangle\right)$ \\\cline{3-4}

      &  $|5\rangle$ &$1^{+-}$&  $|4'\rangle=\frac{1}{\sqrt{2}}\left(|3\rangle-|4\rangle\right)$ \\

      &  $|6\rangle$ &      &  $|6'\rangle=\frac{1}{\sqrt{2}}\left(|5\rangle-|6\rangle\right)$ \\

      &  $|7\rangle$ &      &  $|7\rangle$ \\

      &  $|8\rangle$ &      &  $|8\rangle$ \\\hline

$2^+$ &  $|7\rangle$ &$2^{++}$&  $|7\rangle$ \\

	  &  $|8\rangle$ &      &  $|8\rangle$ \\
\hline\hline
\end{tabular}
\end{table}

\subsubsection{Double-Charm Systems}
\label{fdc}

Similarly as in the previous section, the flavor wave functions of $cc\bar{q}\bar{q}$ systems with proper quantum numbers
can be decomposed as follows:
\begin{itemize}
\item $I=0,\,S=0\,.$
\begin{equation}
\left(T_{cc}^{0,0}\right)^+=\frac{1}{\sqrt{2}}cc\left(-\bar{d}\bar{u}+\bar{u}\bar{d}\right)\,,\label{Tcc1}
\end{equation}
\item $I=0,\,S=2\,.$
\begin{equation}
\left(T_{cc}^{0,2}\right)^{++}=cc\bar{s}\bar{s}\,,\label{Tcc2}
\end{equation}

\item $I=\frac{1}{2},\,S=1\,.$
\begin{align}
\left(T_{cc}^{\frac{1}{2},1}\right)^+&=-\frac{1}{\sqrt{2}}cc\left(\bar{d}\bar{s}\pm \bar{s}\bar{d}\right)\,,\label{Tcc3}\\
\left(T_{cc}^{\frac{1}{2},1}\right)^0&=\frac{1}{\sqrt{2}}cc\left(\bar{u}\bar{s}\pm \bar{s}\bar{u}\right)\,,\label{Tcc4}
\end{align}

\item $I=1,\,S=0\,.$
\begin{align}
&\left(T_{cc}^{1,0}\right)^{++}=cc\bar{d}\bar{d}\,,\label{Tcc5}\\
&\left(T_{cc}^{1,0}\right)^{+}=-\frac{1}{\sqrt{2}}cc\left(\bar{d}\bar{u}+\bar{u}\bar{d}\right)\,,\label{Tcc6}\\
&\left(T_{cc}^{1,0}\right)^{0}=cc\bar{u}\bar{u}\,,\label{Tcc7}
\end{align}
\end{itemize}
here we use $\left(T_{cc}^{I,S}\right)^Q$ to denote $cc\bar{q}\bar{q}$ systems, where $I$, $S$ and $Q$ are isospin, strangeness, and electric charge of the state, respectively.

For the double-charm states, the flavor wave function of diquark must be symmetric since it is $cc$, and the symmetry of antidiquark flavor wave functions is determined by Eqs.~(\ref{Tcc1}\text{--}\ref{Tcc7}). All the configurations of double-charm tetraquark systems and their corresponding quantum numbers are listed in Table~\ref{OCSym}.

\begin{table}[htbp]
\caption{The symmetry configuraitons of double-charm systems.}\label{OCSym}
\renewcommand
\tabcolsep{0.93cm}
\renewcommand{\arraystretch}{2}
\begin{tabular}{ccc}
\hline\hline

$T_{cc}^{I,S}$&$J^{P}$&Configuration\\\hline

$T_{cc}^{0,0}$&$1^{+}$&$|3\rangle$\\

      &       &$|6\rangle$\\\hline

$T_{cc}^{0,2}$&$0^{+}$&$|1\rangle$\\

      &       &$|7\rangle$\\\cline{2-3}

      &$1^{+}$&$|7\rangle$\\\cline{2-3}

      &$2^{+}$&$|7\rangle$\\\hline

$T_{cc}^{\frac{1}{2},1}$&$0^{+}$&$|1\rangle$\\

                &       &$|7\rangle$\\\cline{2-3}

                &$1^{+}$&$|3\rangle$\\

                &       &$|6\rangle$\\

                &       &$|7\rangle$\\\cline{2-3}

                &$2^{+}$&$|7\rangle$\\\hline

$T_{cc}^{1,0}$&$0^{+}$&$|1\rangle$\\

      &       &$|7\rangle$\\\cline{2-3}

      &$1^{+}$&$|7\rangle$\\\cline{2-3}

      &$2^{+}$&$|7\rangle$\\

\hline\hline
\end{tabular}
\end{table}

\subsection{Wave functions}

For the calculation of Hamiltonian matrices, a Gaussian functions expansion method~\cite{Zhang:2007mu,Zhang:2005jz,Liu:2019zuc,GEM} is used here to solve the Schr\"odinger equation of Hamiltonian Eq.~(\ref{H}), where the orbital wave function of the ground $S$-wave tetraquark system can be expanded by a series of Gaussian functions,
\begin{equation}
\Psi(\{\vec{r}_i\})=\prod_{i=1}^{4}\sum_{\mathcal{l}}^{n}C_{i\mathcal{l}}\left(\frac{1}{\pi b_{i\mathcal{l}}^2}\right)^{3/4}\mathrm{exp}\left[-\frac{1}{2b_{i\mathcal{l}}^2}r_i^2\right]\,,
\end{equation}
here $\{b_{i\mathcal{l}}\}$ are harmonic oscillator length parameters and can be related to the frequencies $\{\omega_{\mathcal{l}}\}$ with $1/b_{i\mathcal{l}}^2=m_i\omega_{\mathcal{l}}$, where we have taken the ansatz that $\omega_{\mathcal{l}}$ is independent to the quark mass as in Ref.~\cite{Liu:2019zuc}. Then the orbital wave function can be  simplified to be
\begin{align}
\Psi(\{\vec{r}_i\})=&\sum_{\mathcal{l}}^{n}C_{\mathcal{l}}\prod_{i=1}^{4}\left(\frac{m_i\omega_{\mathcal{l}}}{\pi }\right)^{3/4}\mathrm{exp}\left[-\frac{m_i\omega_{\mathcal{l}}}{2}r_i^2\right]\notag\\
=&\sum_{\mathcal{l}}^{n}C_{\mathcal{l}}\,\psi\left(\omega_{\mathcal{l}},\{\vec{r}_i\}\right)\,,\label{SpatialF}
\end{align}
which is often adopted for the calculations of multiquark systems~\cite{Zhang:2007mu,Zhang:2005jz}.

On the other hand, here we define the Jacobi coordinates by $\{\vec{r}_{i}\}$ as
\begin{eqnarray}
  \vec{\xi}_{1} &=& \vec{r}_{1}-\vec{r}_{2}\,, \\
  \vec{\xi}_{2} &=& \vec{r}_{3}-\vec{r}_{4}\,, \\
  \vec{\xi}_{3} &=& \frac{m_1 \vec{r}_{1}+m_2\vec{r}_{2}}{m_1+m_2}- \frac{m_3 \vec{r}_{3}+m_4\vec{r}_{4}}{m_3+m_4}\,,\\
  \vec{R} &=&  \frac{m_1 \vec{r}_{1}+m_2\vec{r}_{2}+m_3 \vec{r}_{3}+m_4\vec{r}_{4}}{m_1+m_2+m_3+m_4}\,,
\end{eqnarray}
to remove contributions of the motion of the center of mass directly. Where we have enumerated the two quarks by $1,2$ and the two antiquarks by $3,4$, respectively.

In the calculations we define $1/b_{\mathcal{l}}^2\equiv m_q\omega_{\mathcal{l}}$ ($m_{q}$ denotes the constituent mass of the corresponding quark). Following the work of Ref.~\cite{GEM}, the parameters $\{b_{\mathcal{l}}\}$ are set to be geometric series,
\begin{equation}
b_{\mathcal{l}}=b_1a^{\mathcal{l}-1}\hspace{0.8cm}\left(\mathcal{l}=1,2,...,n\right)\,,
\end{equation}
where $n$ is the number of Gaussian functions and $a$ the ratio coefficient, then there are only thee parameters $\{b_{1},b_{n},n\}$ need to be determined.

Using the orbital wave function in Eq.~(\ref{SpatialF}), we compute the diagonal elements of the Hamiltonian matrices by solving the generalized matrix eigenvalue problem:
\begin{equation}
\sum_{\mathcal{l}}^n\sum_{\mathcal{l}'}^nC^i_{\mathcal{l}}\left(H^d_{\mathcal{l}\mathcal{l}'}-E_i^d N_{\mathcal{l}\mathcal{l}'}\right)C^i_{\mathcal{l}'}=0\,,\label{GEP}
\end{equation}
where $i=1\text{--}n$ and
\begin{align}
H^d_{\mathcal{l}\mathcal{l}'}=&\langle\psi\left(\omega_{\mathcal{l}}\right)(FSC)|H_{eff.}|\psi\left(\omega_{\mathcal{l}'}\right)(FSC)\rangle\,,\\
N_{\mathcal{l}\mathcal{l}'}=&\langle\psi\left(\omega_{\mathcal{l}}\right)(FSC)|\psi\left(\omega_{\mathcal{l}'}\right)(FSC)\rangle\,,
\end{align}
here $(FSC)$ stands for flavor$\,\otimes\,$spin$\,\otimes\,$color wave function. One can choose a set of $\{b_{1},b_{n},n\}$ tentatively, then extend and densify the harmonic oscillator length parameters to get a minimum energy $E^d_m$, which should be correlative to the energy of physical state according to the Rayleigh-Ritz variational principle. In present work we take $\{b_{1},b_{n},n\}=\{0.02\,\text{fm},6\,\text{fm},40\}$ for the light quarks, and one can directly obtain the ranges for the strange and charm quarks by replacing the light quark mass by strange and charm quark masses, respectively. Besides, the nondiagonal elements of Hamiltonian matrices can be easily obtained using $\{C_{\mathcal{l}}^m\}$.

\subsection{Model Parameters}

There are sixteen parameters in our model, namely, constituent quark masses $m_n$, $m_s$ and $m_c$; quark confinement strength $b$, QCD effective coupling constants $\alpha_{ij}$ ($i$ and $j$ are quark flavor), and zero point energy $C_0$ for Cornell potential; $g_{nn}$, $g_{ns}$, $g_{nc}$, $g_{sc}$, and $\sigma$ for instanton-induced interaction. All these above model parameters are collected in Table~\ref{ModelParameters}.

\begin{table}[htbp]
\caption{Parameters used in this work.}\label{ModelParameters}
\renewcommand
\tabcolsep{0.25cm}
\renewcommand{\arraystretch}{1.6}
\begin{tabular}{cc}
\hline\hline
$m_n=340\,\text{MeV}$&$m_s=511\,\text{MeV}$\\
$m_c=1674\,\text{MeV}$&\\
\hline
$b=0.39\,\text{GeV}\cdot\text{fm}^{-1}$&$C_0=-296\,\text{MeV}$\\
$\alpha_{nn}=0.600$&$\alpha_{ns}=0.600$\\
$\alpha_{ss}=0.562$&$\alpha_{nc}=0.509$\\
$\alpha_{sc}=0.480$&$\alpha_{cc}=0.450$\\
\hline
$g_{nn}=0.169\times10^{-4}\,\text{MeV}^{-2}$&$g_{ns}=0.107\times10^{-4}\,\text{MeV}^{-2}$\\
$g_{nc}=0.040\times10^{-4}\,\text{MeV}^{-2}$&$g_{sc}=0.031\times10^{-4}\,\text{MeV}^{-2}$\\
$\sigma=485\,\text{MeV}$&\\
\hline\hline
\end{tabular}
\end{table}

Since none of the exotic states is identified as a pure tetraquark state, one cannot take the masses of these exotic states as input to evaluate the model parameters. In our model, the interactions between quarks and antiquarks in tetraquark states are simple superposition of two-body confinements and tow-body residual interactions, which is the same as the description of traditional baryons and mesons in CQM. Therefore, we just extract model parameters from traditional hadron spectra. In this work, eleven mesons and eight baryons are taken as inputs to determine these model parameters.

It is worthy to mention that, if the pseudoscalar mesons whose dominant contents are considered to be pure $q_{f_1}\bar{q}_{f_2}$ with $f_{1}=f_{2}$, as shown in Table~\ref{Tgqqbar}, the instanton-induced interaction effects will vanish, this should lead to the mass relation
$m_{\eta_{c}}=m_{J/\psi}$, it's obviously unreasonable. In fact, as discussed in~\cite{tHooft:1999cta}, the instanton-induced interaction should cause mixing between this kind of states with quantum number $0^{-+}$ naturally, although the mixing between the $c\bar{c}$ pair with the light
quark-antiquark pairs should be very small, it will obviously influence on the mass of $\eta_{c}$. Consequently, to get the correct mass relations, one has to consider the flavor-configuration mixing for $\eta$, $\eta'$ and $\eta_{c}$. However, the mixing angles are unknown and difficult to determine~\cite{tHooft:1999cta}. Thus, these sates $\eta$, $\eta'$ and $\eta_{c}$ are not taken into account for determining the model parameters shown in table~\ref{ModelParameters}. In addition, the effective strong coupling constant between two charm quarks is determined by fitting the mass of $J/\psi$.

With these model parameters shown in Table~\ref{ModelParameters}, the model calculations for the hadron spectra are shown in Table~\ref{FittingResults}.

\begin{table}[htbp]
\caption{Ground state hadron spectrum. The columns denoted by ``M'' are model calculations while the columns denoted by ``PDG'' are  experimental data for the masses of these states in PDG~\cite{PDG}. The units are in MeV.} \label{FittingResults}
\renewcommand
\tabcolsep{0.35cm}
\renewcommand{\arraystretch}{1.6}
\begin{tabular}{cccccc}
\hline\hline
States&M&PDG&States&M&PDG\\
\hline
$\pi   $&$139 $&$138 $&$J/\psi      $&$3097$&$3097$\\
$\rho  $&$775 $&$775 $&$\Sigma_{c}  $&$2447$&$2454$\\
$\omega$&$775 $&$783 $&$\Sigma_{c}^*$&$2518$&$2518$\\
$\phi  $&$1019$&$1019$&$\Xi_{c}     $&$2507$&$2469$\\
$K     $&$499 $&$496 $&$\Xi_{c}'    $&$2554$&$2577$\\
$K^{*} $&$894 $&$895 $&$\Xi_{c}^*   $&$2627$&$2646$\\
$D     $&$1866$&$1867$&$\Omega_{c}  $&$2664$&$2695$\\
$D^{*} $&$2009$&$2009$&$\Omega_{c}^*$&$2737$&$2766$\\
$D_{s} $&$1969$&$1968$&$\Xi_{cc}    $&$3607$&$3621$\\
$D_s^* $&$2111$&$2112$&&&\\
\hline\hline
\end{tabular}
\end{table}

\section{Results and Discussions}
\label{ResultsAndDiscussions}

With the model parameters in Table~\ref{ModelParameters}, one can calculate the mass spectra of the ground $S$-wave hidden- and double-charm tetraquark systems. The numerical results are shown in Tables~\ref{XT}-\ref{TccT}.

As discussed in the Sec.~\ref{Theory}, there are two possible color structures of tetraquark configurations, namely, $\bf{6}_c\otimes\bf{\bar{6}}_c$ and $\bf{\bar{3}}_c\otimes\bf{3}_c$. The results of color factor $\langle\vec{\lambda}^{c}_{i}\cdot\vec{\lambda}^{c}_{j}\rangle$ show that, the Cornell potentials are attractive between each of quarks and antiquarks in tetraquark states with $\bf{\bar{3}}_c\otimes\bf{3}_c$ color structure (configurations $|2\rangle,|3\rangle,|4\rangle,|7\rangle$). However, they are repulsive for those states with $\bf{6}_c\otimes\bf{\bar{6}}_c$ color structure (configurations $|1\rangle,|5\rangle,|6\rangle,|8\rangle$). Intuitively, one might think that the tetraquark states with $\bf{\bar{3}}_c\otimes\bf{3}_c$ structure could be more bound than those states with the $\bf{6}_c\otimes\bf{\bar{6}}_c$, and thus have lower energies.

However, this is not always true. Because, on one hand, a tighter binding is accompanied by a higher quark kinetic energy, on the other hand a color dependent residual interaction, Instanton-induced interaction, is considered in the model, which has a much more complicated interaction structure in tetraquark states compared with traditional meson states.

Consequently, a physical tetraquark state should be composed of the two kinds of components with $\bf{\bar{3}}_c\otimes\bf{3}_c$ and $\bf{6}_c\otimes\bf{\bar{6}}_c$ color structures. In Tables~\ref{XT},~\ref{ZcT},~\ref{ZcsT} and~\ref{TccT}, the numerical results including mixing of the two kinds of components are presented in the last two columns.
Obviously, effects of configurations mixing are important.

In the following subsections, we discuss the results of $X$, $Z_c$, $Z_{cs}$ and double-charm tetraquark states, respectively.

\subsection{$X$ states}

\begin{table*}[htbp]
\caption{The numerical results of ground $S$-wave $X$ states with $I^G = 0^+$. First and second columns are the quark content of each state and the corresponding $J^{PC}$ quantum numbers, the fourth column are the results of single configuration calculations, fifth and sixth columns are the model results after considering configuration mixing, but not considering $cn\bar{c}\bar{n}$ and $cs\bar{c}\bar{s}$ mixing.}\label{XT}
\renewcommand
\tabcolsep{0.42cm}
\renewcommand{\arraystretch}{1.5}
\begin{tabular}{cccccc}
\hline\hline

 & &\multicolumn{2}{c}{Single configuration}&\multicolumn{2}{c}{Configurations mixing}\\
 \hline

Quark content                      &$J^{PC}                 $ &Config.     &Energies\,(MeV)&Energies\,(MeV)&Mixing coefficients\\
\hline
\multirow{4}{*}{$c\bar{c}n\bar{n}$}&\multirow{4}{*}{$0^{++}$}&$|1 \rangle$&$4161.55    $&$4039.34    $&$(-0.64,\hspace{0.15cm}-0.43,\hspace{0.15cm} 0.00,\hspace{0.15cm}-0.64)$\\

                                   &                              &$|2 \rangle$&$4131.90    $&$4115.93    $&$( 0.00,\hspace{0.15cm}-0.75,\hspace{0.15cm}-0.43,\hspace{0.15cm} 0.50)$\\

                                   &                              &$|7 \rangle$&$4212.69    $&$4231.32    $&$( 0.28,\hspace{0.15cm}-0.46,\hspace{0.15cm} 0.84,\hspace{0.15cm} 0.02)$\\

                                   &                              &$|8 \rangle$&$4131.28    $&$4250.83    $&$( 0.71,\hspace{0.15cm}-0.20,\hspace{0.15cm}-0.34,\hspace{0.15cm}-0.59)$\\
\hline
\multirow{2}{*}{$c\bar{c}n\bar{n}$}&\multirow{2}{*}{$1^{++}$}&$|3'\rangle$&$4151.06    $&$4052.78    $&$(0.51,\hspace{0.15cm}-0.86)$\\

                                   &                              &$|5'\rangle$&$4088.23    $&$4186.52    $&$(-0.86,\hspace{0.15cm}-0.51)$\\
                                   \hline
\multirow{4}{*}{$c\bar{c}n\bar{n}$}&\multirow{4}{*}{$1^{+-}$}&$|4'\rangle$&$4193.21    $&$4053.08    $&$(-0.46,\hspace{0.15cm} 0.51,\hspace{0.15cm}-0.25,\hspace{0.15cm} 0.68)$\\

								   &                              &$|6'\rangle$&$4190.93    $&$4186.87    $&$(-0.26,\hspace{0.15cm} 0.41,\hspace{0.15cm}-0.55,\hspace{0.15cm}-0.68)$\\

								   &                              &$|7 \rangle$&$4211.40    $&$4228.55    $&$( 0.75,\hspace{0.15cm} 0.05,\hspace{0.15cm}-0.61,\hspace{0.15cm} 0.24)$\\

								   &                              &$|8 \rangle$&$4127.50    $&$4254.52    $&$(0.40,\hspace{0.15cm}0.76,\hspace{0.15cm} 0.51,\hspace{0.15cm}-0.11)$\\
								   \hline
\multirow{2}{*}{$c\bar{c}n\bar{n}$}&\multirow{2}{*}{$2^{++}$}&$|7 \rangle$&$4208.78    $&$4055.73    $&$(-0.54,\hspace{0.15cm}-0.84)$\\

                                   &                              &$|8 \rangle$&$4119.53    $&$4272.57    $&$(-0.84,\hspace{0.15cm} 0.54)$\\	
                                   \hline
\multirow{4}{*}{$c\bar{c}s\bar{s}$}&\multirow{4}{*}{$0^{++}$}&$|1 \rangle$&$4341.92    $&$4203.76    $&$(-0.24,\hspace{0.15cm}-0.37,\hspace{0.15cm}-0.14,\hspace{0.15cm} 0.89)$\\

                                   &                              &$|2 \rangle$&$4293.15    $&$4270.15    $&$( 0.61,\hspace{0.15cm}-0.77,\hspace{0.15cm} 0.10,\hspace{0.15cm}-0.14)$\\

                                   &                              &$|7 \rangle$&$4361.49    $&$4340.17    $&$( 0.34,\hspace{0.15cm} 0.32,\hspace{0.15cm} 0.81,\hspace{0.15cm} 0.35)$\\

                                   &                              &$|8 \rangle$&$4237.31    $&$4419.79    $&$(-0.67,\hspace{0.15cm}-0.41,\hspace{0.15cm} 0.56,\hspace{0.15cm}-0.27)$\\
\hline
\multirow{2}{*}{$c\bar{c}s\bar{s}$}&\multirow{2}{*}{$1^{++}$}&$|3'\rangle$&$4328.42    $&$4284.16    $&$(-0.39,\hspace{0.15cm}-0.92)$\\

                                   &                              &$|5'\rangle$&$4292.14    $&$4336.39    $&$(-0.92,\hspace{0.15cm} 0.39)$\\
\hline
\multirow{4}{*}{$c\bar{c}s\bar{s}$}&\multirow{4}{*}{$1^{+-}$}&$|4'\rangle$&$4350.70    $&$4265.06    $&$( 0.00,\hspace{0.15cm} 0.00,\hspace{0.15cm}-0.17,\hspace{0.15cm} 0.98)$\\

                                   &                              &$|6'\rangle$&$4347.16    $&$4283.64    $&$( 0.70,\hspace{0.15cm}-0.72,\hspace{0.15cm} 0.00,\hspace{0.15cm} 0.00)$\\

                                   &                              &$|7 \rangle$&$4372.05    $&$4375.42    $&$( 0.00,\hspace{0.15cm} 0.00,\hspace{0.15cm} 0.98,\hspace{0.15cm} 0.17)$\\

                                   &                              &$|8 \rangle$&$4268.42    $&$4414.23    $&$(0.72,\hspace{0.15cm}0.70,\hspace{0.15cm} 0.00,\hspace{0.15cm} 0.00)$\\
\hline
\multirow{2}{*}{$c\bar{c}s\bar{s}$}&\multirow{2}{*}{$2^{++}$}&$|7 \rangle$&$4392.52    $&$4286.55    $&$( 0.52,\hspace{0.15cm}-0.85)$\\

                                   &                              &$|8 \rangle$&$4325.86    $&$4431.83    $&$(-0.85,\hspace{0.15cm}-0.52)$\\
\hline\hline                                                                      							
\end{tabular}
\end{table*}

In this subsection, we present the numerical results for spectrum of the ground $S$-wave $cn\bar{c}\bar{n}$ and $cs\bar{c}\bar{s}$ states, those we call $X$ states.

Similar to $\eta\text{--}\eta'$ mixing, there are transitions between $cn\bar{c}\bar{n}$ and $cs\bar{c}\bar{s}$ quark configurations, which can be naturally caused by the instanton-induced interaction.
Consequently, we will discuss the results within two models: one takes into account only the single configurations
$cn\bar{c}\bar{n}$ or $cs\bar{c}\bar{s}$, respectively; while the other one includes the mixing of these configurations. The numerical results of energies for the $I^{G}=0^+$ $X$ states are given in Tables~\ref{XT}~and~\ref{XmT}, respectively, where the masses and configurations mixing coefficients for the obtained tetraquark states are shown explicitly.

\begin{figure*}[htbp]
	\centering
	\includegraphics[scale=0.6]{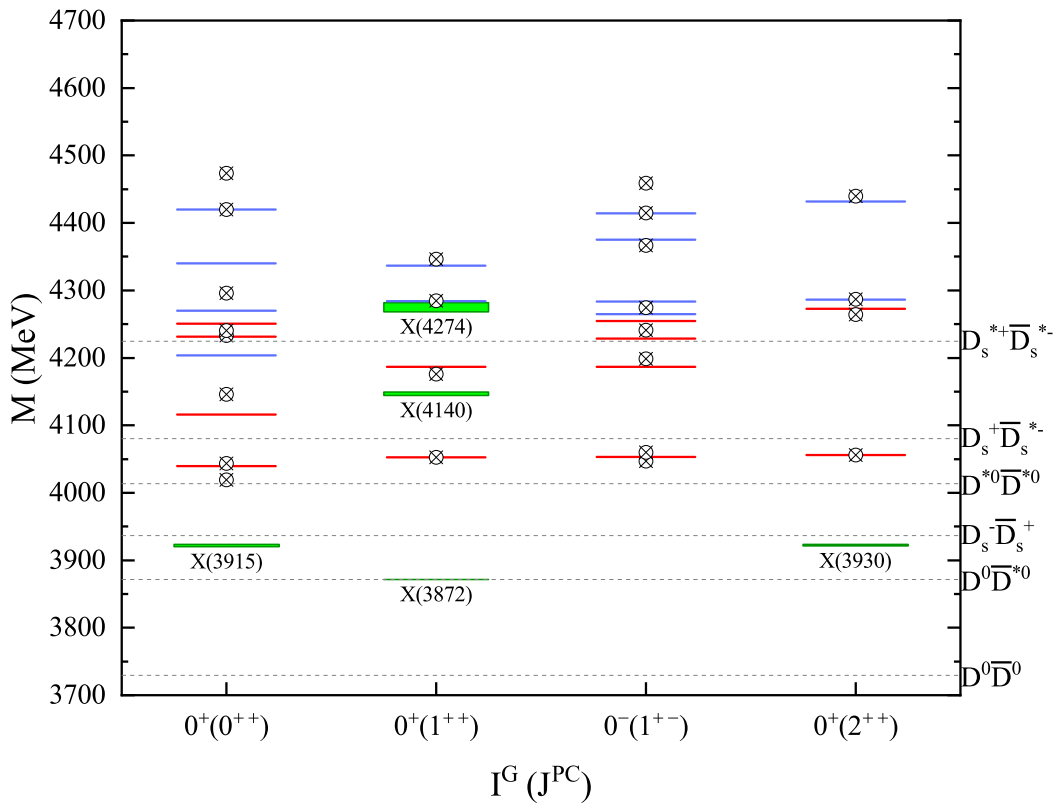}
	\caption{Mass spectrum of ground $S$-wave $X$ states. Red solid lines denote $cn\bar{c}\bar{n}$ system and blue lines denote $cs\bar{c}\bar{s}$ system, hollow circles with crosses denote results after considering $cn\bar{c}\bar{n}$ and $cs\bar{c}\bar{s}$ mixing, gray dotted lines are corresponding $S$-wave charm meson pair thresholds, the green rectangles represent $X$ states experimental masses taken from PDG~\cite{PDG}, the rectangle width stand their masses uncertainties.}\label{XF}
\end{figure*}

Meanwhile, we depict the obtained mass spectrum of $X$ states in Fig.~\ref{XF} comparing with the experimental measurements, where red solid lines denote $cn\bar{c}\bar{n}$ system and blue lines denote $cs\bar{c}\bar{s}$ system, hollow circles with crosses denote results after considering $cn\bar{c}\bar{n}$ and $cs\bar{c}\bar{s}$ mixing, gray dotted lines are corresponding $S$-wave charm meson pair thresholds, finally, the green rectangles represent experimental masses of these $X$ states taken from PDG~\cite{PDG}, the rectangle widths stand their masses uncertainties.

Experimentally, in 2003, the $X(3872)$ state was observed by Belle collaboration~\cite{Belle 2003} and its quantum number was determined to be $J^{PC}=1^{++}$~\cite{LHCb 2013}. Explicitly, its mass and width are: $M=3871.65 \pm0.06$ and $\Gamma=1.19\pm0.01\,\text{MeV}$~\cite{PDG}. Its mass is extremely close to the $D^{0}\bar{D}^{*0}$ mass threshold, which indicates that $X(3872)$ may be explained as a $D^{0}\bar{D}^{*0}$ hadronic molecule. In present work, the lowest energy of $cn\bar{c}\bar{n}$ states with quantum numbers $I^{G}(J^{PC})=0^+(1^{++})$ is about $4053\,\text{MeV}$, which is much higher than the mass of $X(3872)$. Consequently, our result may support that the $X(3872)$ is mostly dominated by a hadronic molecular state.

The $X(3930)$ and $X(3915)$ states are observed through two photon fusion processes by Belle in 2005~\cite{Belle 2006} and 2010~\cite{Belle 2010}, respectively. They were suggested to be good candidates of the $2^3P_0$ and $2^3P_2$ charmonium states, $\chi_{c0}(2P)$ and $\chi_{c2}(2P)$, respectively~\cite{Belle 2006,Belle 2010,BaBar 2012}.
From our results shown in Table~\ref{XT} and Fig.~\ref{XF}, it is found that our results of the states with quantum numbers $I^{G}(J^{PC}) = 0^+(0^{++})$ and $I^{G}(J^{PC}) = 0^+(2^{++})$ are all higher than the masses of $X(3930)$ and $X(3915)$. Thus $X(3930)$ and $X(3915)$ cannot be described by the compact tetraquark states.

The CDF collaboration reported the discoveries of $X(4140)$ and $X(4274)$ in the $J/\psi\phi$ invariant mass distribution of the process $B^+\rightarrow J/\psi\phi K^+$ in 2009~\cite{CDF 2009} and 2011~\cite{CDF 2011}, respectively. Later, LHCb collaboration confirmed their existences and determined their spin-parity quantum numbers to be $J^{P}=1^{+}$~\cite{LHCb 2017, LHCb PRD 2017, LHCb 2021}. Masses and widths of $X(4140)$ and $X(4274)$ are~\cite{PDG}
\begin{align}
&(M_{X(4140)}=4146.8\pm2.4\,,\Gamma_{X(4140)}=22^{+8}_{-7})\,\text{MeV}\,,\notag\\
&(M_{X(4274)}=4274^{+8}_{-6}\,,\Gamma_{X(4274)}=49\pm12)\,\text{MeV}\,,\notag
\end{align}
respectively.
The next lowest energy of the $cn\bar{c}\bar{n}$ states with quantum number $I^{G}(J^{PC})=0^+(1^{++})$ in present model is around $4187\,\text{MeV}$, very close to the mass of $X(4140)$ state, while the lowest energy of the $cs\bar{c}\bar{s}$ states with quantum number $I^{G}(J^{PC})=0^+(1^{++})$ in present model is about $4284\,\text{MeV}$, very close to the mass of $X(4274)$ state. Thus $X(4140)$ and $X(4274)$ can be interpreted as compact $c\bar{c}n\bar{n}$ and $c\bar{c}s\bar{s}$ tetraquark states with quantum number $I^{G}(J^{PC})=0^+(1^{++})$, respectively.

The results by considering $cn\bar{c}\bar{n}$ and $cs\bar{c}\bar{s}$ mixing are also given in Fig.~\ref{XF}, which are denoted by hollow circles with crosses.
It can be found that configurations mixing effects in the tetraquark states with quantum numbers $J^{PC}=0^{++}$ are significant, while those
in the $J^{PC}=1^{++}$ tetraquark states are tiny. The results of four of the obtained $X$ states with quantum number $J^{PC}=1^{++}$, considering $cn\bar{c}\bar{n}$ and $cs\bar{c}\bar{s}$ mixing, are shown in Table~\ref{XmT} with explicit probabilities of the admixtures. One can find that every state is dominated by a single configuration.

\begin{table}[htbp]
\caption{The results of the $X$ states with quantum numbers $J^{PC}=1^{++}$ by considering $cn\bar{c}\bar{n}$ and $cs\bar{c}\bar{s}$ mixing.}
\label{XmT}
\renewcommand
\tabcolsep{0.25cm}
\renewcommand{\arraystretch}{1.5}
\begin{tabular}{ccccc}
\hline\hline
\multicolumn{3}{c}{Model values}&\multicolumn{2}{c}{Experimental values}\\
M\,(MeV)&$c\bar{c}n\bar{n}$&$c\bar{c}s\bar{s}$&States&M\,(MeV)\\
\hline
$4052.72    $&$99.98\%         $&$0.02\%          $&---&---\\

$4176.07    $&$93.70\%         $&$6.30\%          $&$X(4140)$&$4146.8\pm2.4$\\

$4284.76    $&$0.46\%          $&$99.54\%         $&$X(4274)$&$4274^{+8}_{-6}$\\

$4346.29    $&$5.86\%          $&$94.14\%         $&---&---\\
\hline\hline
\end{tabular}
\end{table}

Finally, as shown in Fig.~\ref{XF}, one may note that most of the energies for the obtained $X$ states in present work are near but higher than thresholds of corresponding $S-$wave meson-meson channels, with only a few exceptions including the states shown in Table~\ref{XmT}. This may indicate that most of the presently obtained $X$ states with compact tetraquark structure cannot form considerable components in the physical meson exotics.

\subsection{$Z_c$ and $Z_{cs}$ states}

The numerical results for $S$-wave $cn\bar{c}\bar{n}$ and $cn\bar{c}\bar{s}$ systems are collected in Table.~\ref{ZcT} and Table.~\ref{ZcsT}, respectively. In these tables, the first and second columns give the quark content of each state and their corresponding quantum numbers, the fourth column shows the numerical results considering only single configurations, while numbers in the fifth column are the results obtained by including the configurations mixing effects, and the explicit probability amplitudes for the mixed configurations are shown in the last column.

\begin{table*}[htbp]
\caption{As in table~\ref{XT} but for the case of $Z_c$ states.}\label{ZcT}
\renewcommand
\tabcolsep{0.42cm}
\renewcommand{\arraystretch}{1.5}
\begin{tabular}{cccccc}
\hline\hline

  & &\multicolumn{2}{c}{Single configuration}&\multicolumn{2}{c}{Configurations mixing}\\
 \hline

Quark content                      &$I^G(J^{PC})                 $&Config.     &Energies\,(MeV)&Energies\,(MeV)&Mixing coefficients\\
\hline\hline		
\multirow{4}{*}{$c\bar{c}n\bar{n}$}&\multirow{4}{*}{$1^-(0^{++})$}&$|1 \rangle$&$4079.46      $&$3777.23      $&$( 0.23,\hspace{0.15cm}0.42,\hspace{0.15cm} 0.30,\hspace{0.15cm} 0.82)$\\
		
                                   &                              &$|2 \rangle$&$4048.33      $&$4018.55      $&$(-0.76,\hspace{0.15cm}-0.17,\hspace{0.15cm}-0.42,\hspace{0.15cm} 0.46)$\\

                                   &                              &$|7 \rangle$&$4093.05      $&$4076.61      $&$(-0.04,\hspace{0.15cm}-0.78,\hspace{0.15cm} 0.60,\hspace{0.15cm} 0.19)$\\

                                   &                              &$|8 \rangle$&$3872.81      $&$4221.26      $&$(-0.60,\hspace{0.15cm} 0.43,\hspace{0.15cm} 0.61,\hspace{0.15cm}-0.28)$\\
                                   \hline
\multirow{2}{*}{$c\bar{c}n\bar{n}$}&\multirow{2}{*}{$1^-(1^{++})$}&$|3'\rangle$&$4099.70      $&$4052.82      $&$(-0.40,\hspace{0.15cm}-0.91)$\\

 								   &                              &$|5'\rangle$&$4061.95      $&$4108.84      $&$(-0.91,\hspace{0.15cm}0.40)$\\
 \hline
\multirow{4}{*}{$c\bar{c}n\bar{n}$}&\multirow{4}{*}{$1^+(1^{+-})$}&$|4'\rangle$&$4095.30      $&$3814.57      $&$(-0.43,\hspace{0.15cm}-0.53,\hspace{0.15cm}-0.31,\hspace{0.15cm} 0.66)$\\

                                   &                              &$|6'\rangle$&$4050.89      $&$4050.16      $&$(0.48,\hspace{0.15cm}0.54,\hspace{0.15cm}-0.14,\hspace{0.15cm} 0.68)$\\

                                   &                              &$|7 \rangle$&$4118.37      $&$4149.29      $&$(-0.28,\hspace{0.15cm} 0.09,\hspace{0.15cm} 0.90,\hspace{0.15cm} 0.32)$\\

                                   &                              &$|8 \rangle$&$3959.10      $&$4209.65      $&$( 0.71,\hspace{0.15cm}-0.65,\hspace{0.15cm} 0.26,\hspace{0.15cm} 0.07)$\\
\hline
\multirow{2}{*}{$c\bar{c}n\bar{n}$}&\multirow{2}{*}{$1^-(2^{++})$}&$|7 \rangle$&$4164.70      $&$4052.98      $&$(-0.52,\hspace{0.15cm}-0.85)$\\

                                   &                              &$|8 \rangle$&$4095.25      $&$4206.97      $&$(-0.85,\hspace{0.15cm}0.52)$\\
\hline\hline
\end{tabular}
\end{table*}

\begin{table*}[htbp]
\caption{As in table~\ref{XT} but for the case of $Z_{cs}$ states.}\label{ZcsT}
\renewcommand
\tabcolsep{0.3cm}
\renewcommand{\arraystretch}{1.5}
\begin{tabular}{cccccc}
\hline\hline

 & &\multicolumn{2}{c}{Single configuration}&\multicolumn{2}{c}{Configurations mixing}\\
 \hline
		
Quark content                      &$I(J^{P})                       $&Config.     &Energies\,(MeV)&Energies\,(MeV)&Mixing coefficients\\
\hline\hline		
\multirow{4}{*}{$c\bar{c}n\bar{s}$}&\multirow{4}{*}{$\frac12(0^{+})$}&$|1 \rangle$&$4198.04      $&$3951.72      $&$( 0.17,\hspace{0.15cm}0.45,\hspace{0.15cm}-0.24,\hspace{0.15cm} 0.84)$\\
		
		                           &                                 &$|2 \rangle$&$4158.90      $&$4129.62      $&$(-0.76,\hspace{0.15cm}-0.23,\hspace{0.15cm} 0.45,\hspace{0.15cm} 0.41)$\\
		
		                           &                                 &$|7 \rangle$&$4210.36      $&$4186.38      $&$( 0.00,\hspace{0.15cm}-0.76,\hspace{0.15cm}-0.61,\hspace{0.15cm} 0.23)$\\
		
		                           &                                 &$|8 \rangle$&$4019.17      $&$4318.75      $&$(-0.62,\hspace{0.15cm} 0.41,\hspace{0.15cm}-0.61,\hspace{0.15cm}-0.26)$\\
		\hline
\multirow{6}{*}{$c\bar{c}n\bar{s}$}&\multirow{6}{*}{$\frac12(1^{+})$}&$|3 \rangle$&$4205.97      $&$3988.72      $&$(-0.31,\hspace{0.15cm}-0.31,\hspace{0.15cm}0.37,\hspace{0.15cm}-0.37,\hspace{0.15cm}0.29,\hspace{0.15cm}0.68)$\\
		
		                           &                                 &$|4 \rangle$&$4209.70      $&$4159.09      $&$(0.34,\hspace{0.15cm}0.34,\hspace{0.15cm}-0.28,\hspace{0.15cm}0.48,\hspace{0.15cm}0.10,\hspace{0.15cm}0.67)$\\
		
                                   &                                 &$|5 \rangle$&$4176.11      $&$4165.97      $&$(0.31,\hspace{0.15cm}-0.21,\hspace{0.15cm}-0.70,\hspace{0.15cm}-0.60,\hspace{0.15cm}0.03,\hspace{0.15cm}0.09)$\\
		
		                           &                                 &$|6 \rangle$&$4174.25      $&$4212.73      $&$(0.64,\hspace{0.15cm}-0.67,\hspace{0.15cm}0.28,\hspace{0.15cm}0.23,\hspace{0.15cm}-0.05,\hspace{0.15cm}-0.02)$\\
		
		                           &                                 &$|7 \rangle$&$4231.97      $&$4253.47      $&$(-0.20,\hspace{0.15cm}-0.16,\hspace{0.15cm}-0.07,\hspace{0.15cm}0.01,\hspace{0.15cm}-0.92,\hspace{0.15cm}0.28)$\\
		
		                           &                                 &$|8 \rangle$&$4089.37      $&$4307.39      $&$(-0.49,\hspace{0.15cm}-0.52,\hspace{0.15cm}-0.47,\hspace{0.15cm}0.47,\hspace{0.15cm}0.22,\hspace{0.15cm}-0.05)$\\
		\hline
\multirow{2}{*}{$c\bar{c}n\bar{s}$}&\multirow{2}{*}{$\frac12(2^{+})$}&$|7 \rangle$&$4272.18      $&$4167.21      $&$(0.52,\hspace{0.15cm}-0.85)$\\
		
	                           	   &                                 &$|8 \rangle$&$4205.90      $&$4310.87      $&$(-0.85,\hspace{0.15cm}-0.52)$\\
\hline\hline
	\end{tabular}
\end{table*}

\begin{figure*}[htbp]
	\centering
	\includegraphics[scale=0.6]{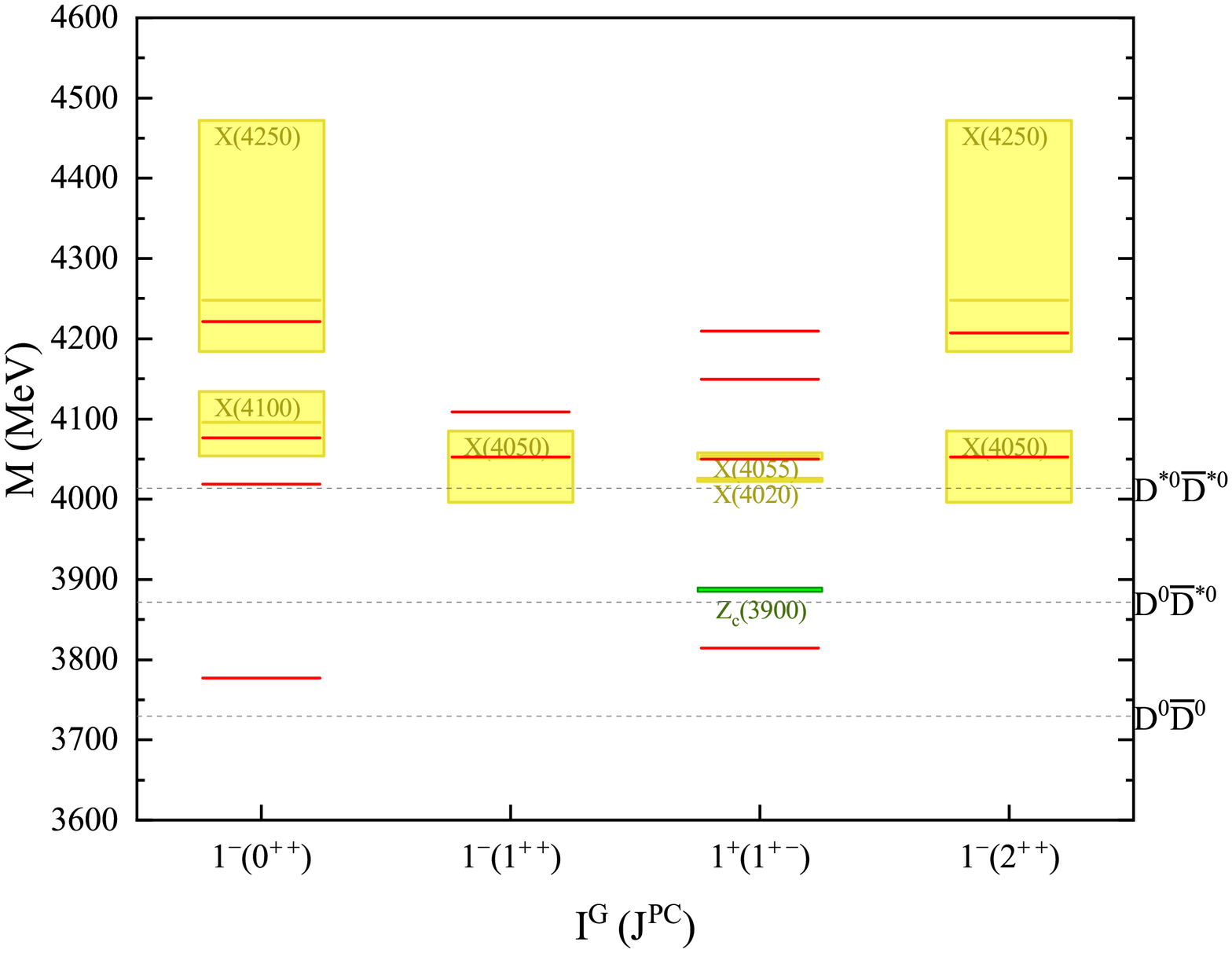}
	\caption{Mass spectra of ground $S$-wave $Z_{c}$ states. Red solid lines denote the model results. Gray dotted lines are corresponding $S$-wave charm meson pair thresholds. The green rectangle represents experimental mass of $Z_c(3900)$ with definite quantum number, the rectangle width is its mass uncertainty. The yellow rectangles represents the electric charmonium-like states, observed in experiments near 4.1~GeV, whose spin-parity quantum numbers have not been pinned down yet, the corresponding quantum numbers in figure are suggested by model calculation, the rectangle width stand their masses uncertainties, while the deep yellow solid lines in rectangles are central values of their experimental masses. All the experimental values shown in this figure are taken from PDG~\cite{PDG}.}\label{ZcF}
\end{figure*}

\begin{figure*}[htbp]
	\centering
	\includegraphics[scale=0.6]{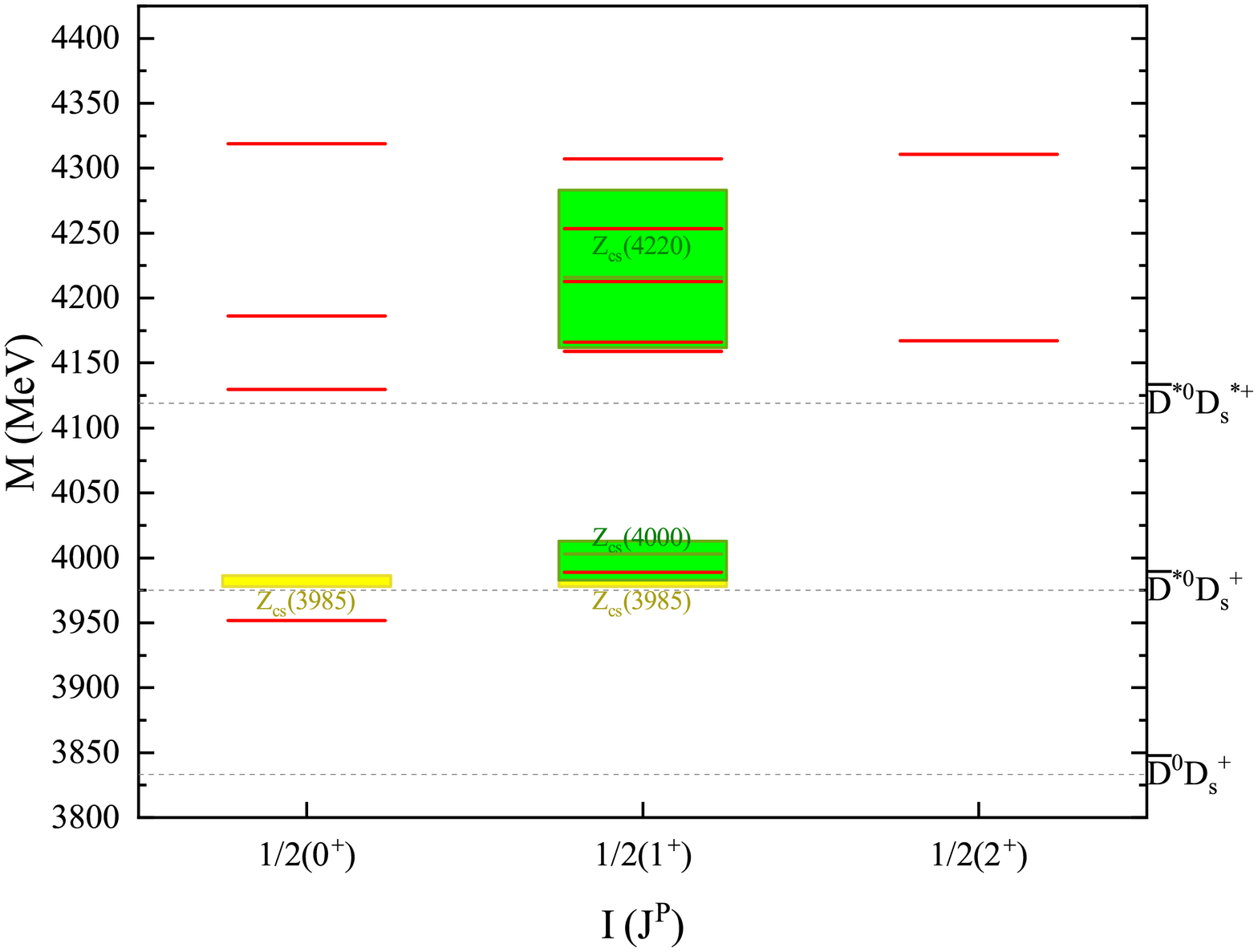}
	\caption{Mass spectra of ground $S$-wave $Z_{cs}$ states. Red solid lines denote the model results. Gray dotted lines are corresponding $S$-wave charm meson pair thresholds. The green rectangles represent the experimental masses of $Z_{cs}(4000)$ and $Z_{cs}(4220)$ with definite quantum numbers~\cite{LHCb 2021}, the rectangle width stand their masses uncertainties and the deep green solid lines in the rectangles are central values of their masses. The yellow rectangles represent the experimental mass of $Z_{cs}(3985)$, the rectangle width stand its mass uncertainty. Spin-parity quantum numbers of $Z_{cs}(3985)$ in figure are suggested by model calculation.}\label{ZcsF}
\end{figure*}

In addition, we also depict the mass spectra of $S$-wave $cn\bar{c}\bar{n}$ and $cn\bar{c}\bar{s}$ tetraquark states in Figs.~\ref{ZcF} and \ref{ZcsF} comparing with the experimental measurements~\cite{PDG,LHCb 2021,BESIII 2021}, where the red solid lines are results obtained in present work, while the green and yellow rectangles represent the states with and without definite quantum numbers in experiments, respectively, with the rectangle widths indicating their masses uncertainties, and the gray dotted lines denote the corresponding $S$-wave charm meson pair thresholds.

In 2013, BESIII collaboration reported the electric charmoniumlike state $Z_{c}(3900)^{\pm}$ in the $\pi^{\pm}J/\psi$ invariant mass spectrum of the process $e^+e^-\rightarrow\pi^+\pi^-J/\psi$~\cite{BESIII 2013}. And the same structure was also observed in the same process by Belle collaboration~\cite{Belle 2013}. Later on, $Z_c(3885)^{\pm}$ was discovered in the $(D\bar{D}^*)^{\pm}$ invariant mass spectrum of the process $e^+e^-\rightarrow\pi^{\mp}(D\bar{D}^*)^{\pm}$ by BESIII collaboration, which should have a very similar mass with the $Z_{c}(3900)^{\pm}$ state~\cite{BESIII 2014}. Since they have the same spin–parity $J^{P}=1^+$, similar mass and width, $Z_{c}(3900)$ and $Z_c(3885)$  are probably the same state, and mainly couple dominantly to the $D\bar{D}^*$ channel~\cite{BESIII 2014}. Thus we will not distinguish the $Z_c(3900)$ and $Z_c(3885)$ in the following discussions.

The average mass and width of $Z_{c}(3900)$ from PDG~\cite{PDG} are: $M_{Z_c(3900)}=3887.1\pm2.6$ and $\Gamma_{Z_c(3900)}=28.4\pm2.6\,\text{MeV}$. In present model, the lowest mass of ground $S$-wave $cn\bar{c}\bar{n}$ states with quantum numbers $I^G(J^{PC})=1^{+}(1^{+-})$, considering configuration mixing, is about $3815\,\text{MeV}$. This value is lower than, but not far from, the experimental mass of $Z_c(3900)$. This may indicate that the compact $c\bar{c}n\bar{n}$ tetraquark components may take notable probability in the $Z_{c}(3900)$ state.

The $Z_c(4025)$ state was observed in $D\bar{D}^*$ final state in Ref.~\cite{BESIII 2014 Zc4025} and, meanwhile, a similar structure $Z_c(4020)$ was observed in $\pi h_c$ final state by BESIII collaboration~\cite{BESIII 2013 Zc4020}. At present, the $Z_c(4020)$ and $Z_c(4025)$ states are denoted as an identical state $X(4020)$ in PDG~\cite{PDG}. Its average mass and width are $M_{Z_c(4020)} =4024.1\pm1.9,\,\Gamma_{Z_c(4020)}=13\pm5\,\text{MeV}$, and its quantum number are probably $J^{PC}=1^{+-}$~\cite{HeJ 2013}. In present model, the next-lowest mass of ground $S$-wave $cn\bar{c}\bar{n}$ state with quantum number $I^G(J^{PC})=1^{+}(1^{+-})$, considering configuration mixing, is about $4050\,\text{MeV}$, which is very close to the mass of $X(4020)$. Therefore, one may expect that the $c\bar{c}n\bar{n}$ tetraquark configuration obtained here could be a component in the $X(4020)$ state.

\begin{table}[htbp]
\caption{The comparison of model results with experimental masses for $Z_c$ states. Fourth, fifth and sixth columns collect the experimental data of $Z_c$ states near $4.1\,\text{GeV}$, and first two columns collect the model results in which model masses are close to the experimental masses.}\label{ZcCT}
	\renewcommand
	\tabcolsep{0.18cm}
	\renewcommand{\arraystretch}{1.5}
	\begin{tabular}{ccccc}
\hline\hline
\multicolumn{2}{c}{Present results}&\multicolumn{3}{c}{Experimental data}\\
$I^G(J^{PC})$&  M\,(MeV) &   States  &$I^G(J^{PC})$&    M\,(MeV)   \\
\hline
$1^+(1^{+-})$&$3814.57$&$Z_c(3900)$&$1^+(1^{+-})$&$3887.1\pm2.6$\\
\hline
$1^+(1^{+-})$&$4050.16$&$X(4020)  $&$1^+(?^{?-})$&$4024.1\pm1.9$\\
\hline
$1^-(1^{++})$&$4052.82$&\multirow{2}{*}{$X(4050)$}&\multirow{2}{*}{$1^-(?^{?+})$}&\multirow{2}{*}{$4051\pm14^{+20}_{-41}$}\\

$1^-(2^{++})$&$4052.98$&                        &                 &\\
\hline
$1^+(1^{+-})$&$4050.16$&$X(4055)  $&$1^+(?^{?-})$&$4054\pm3\pm1$\\
\hline
$1^-(0^{++})$&$4076.61$&$X(4100)$&$1^-(?^{??})$&$4096\pm20^{+18}_{-22}$\\
\hline
$1^-(0^{++})$&$4221.26$&\multirow{2}{*}{$X(4250)$}&\multirow{2}{*}{$1^-(?^{?+})$}&\multirow{2}{*}{$4248^{+44+180}_{-29-35}$}\\

$1^-(2^{++})$&$4206.97$&   &   &   \\
\hline\hline
	\end{tabular}
\end{table}

Besides, there are several other electric charmonium-like particles observed experimentally, which need to be further confirmed.
Here we also compare the present numerical results with these experimental data,
as shown in the Table~\ref{ZcCT}, the last three columns collect the experimental data of $Z_c$ states near $4.1\,\text{GeV}$, and the first two columns are the presently obtained numerical results those are close to the experimental data.

The $X(4050)$ and $X(4250)$ states were reported by Belle collaboration in the $\pi^+\chi_{c1}$ invariant mass distribution of the process $\bar{B}^0\rightarrow K^-\pi^+\chi_{c1}$~\cite{Belle 2008}. In present model, we get two states with masses very close two $X(4050)$, whose quantum numbers are $I^G(J^{PC})=1^{-}(1^{++})$ and $I^G(J^{PC})=1^{-}(2^{++})$, respectively. Meanwhile, masses of two obtained states
with the quantum numbers $I^G(J^{PC})=1^{-}(0^{++})$ and $I^G(J^{PC})=1^{-}(2^{++})$, respectively, are a little bit lower than $X(4250)$.
Finally, two obtained states, whose quantum numbers are $I^G(J^{PC})=1^{+}(1^{+-})$ and $1^{-}(0^{++})$, respectively,
are very close to the $X(4055)^{\pm}$ observed in $\pi^{\pm}\psi$ final state through process $e^+e^-\rightarrow\pi^+\pi^-\psi(2S)$ by Belle~\cite{Belle 2015}, and $X(4100)$ observed in the $\eta_c\pi^-$ invariant mass spectrum in $B^0\rightarrow\eta_cK^+\pi^-$ decay by LHCb collaboration~\cite{LHCb 2018}. We look forward to further examinations on these results by future experiments.

Now we turn to the $Z_{cs}$ states. The $Z_{cs}(3985)$ was the first observed charmonium-like state containing a strange quark, it was discovered in the $D_{s}^{-}D^{*0}(D_{s}^{*-}D^0)$ channel by BESIII very recently~\cite{BESIII 2021}. The mass and width of $Z_{cs}(3985)$ are $M_{Z_{cs}(3985)}=3982.5^{+1.8}_{-2.6}\pm{2.1}$ and $\Gamma_{Z_{cs}(3985)}=12.8^{+5.3}_{-4.4}\pm3.0\,\text{MeV}$. Its mass is very close to the $D_{s}^{-}D^{*0}(D_{s}^{*-}D^0)$ mass threshold. Recently, an theoretical analysis shows that the $Z_{cs}(3985)$ cannot be a pure $D_{s}^{-}D^{*0}(D_{s}^{*-}D^0)$ molecule~\cite{ChenR 2021}. And in Ref.~\cite{Guo 2021}, it has been shown that the two charm-meson molecular components make up about 40 percent of the physical $Z_{cs}(3985)$ state.

In addition, the $Z_{cs}(3985)$ state has also been studied within constituent quark models~\cite{JinX 2021,YangG 2021}. For example, in Ref.~\cite{JinX 2021}, the diquark-antidiquark $[cs][\bar{c}\bar{n}]$ configurations have been investigated,
and it's found that there would be resonance states falling in the mass range of $3916.5\sim3964.6$\,MeV with quantum numbers $J^P=0^+$, and $4008.8\sim4091.2$\,MeV with $J^P=1^+$. Thus, one may conclude that a compact tetraquark state with $J^P=0^+$ or $J^P=1^+$
may take considerable probability in $Z_{cs}(3985)$. And in a very recent paper~\cite{Ikeno:2021mcb}, it's shown that the $Z_{cs}$
state may correspond to a cusp structure.

In present model, the lowest mass of ground $S$-wave $cn\bar{c}\bar{s}$ states with quantum number $I(J^{P})=1/2(0^+)$ and $I(J^{P})=1/2(1^+)$ are about $3952\,\text{MeV}$ and $3989\,\text{MeV}$, respectively. Both of them are close to the experimental mass of $Z_{cs}(3985)$. Therefore, based on the above discussions, our results also support that the $Z_{cs}(3985)$ may be not a
pure hadronic molecule, and the compact tetraquark component should take notable probability in it.

Besides, LHCb collaboration performed an analysis of the $B^+\rightarrow J/\psi\phi K^+$ dacay and reported two $Z_{cs}$ particles, $Z_{cs}(4000)$ and $Z_{cs}(4220)$ in the invariant $J/\psi K^+$ mass distributions~\cite{LHCb 2021}. The spin-parity of $Z_{cs}(4000)$ is $J^{P}=1^+$ and $Z_{cs}(4220)$ is $J^{P}=1^+$ or $J^{P}=1^-$. Their fitting masses and widths are
\begin{align}
&M_{Z_{cs}(4000)}=4003\pm6^{+4}_{-14}\,\text{MeV}\,,\notag\\
&\Gamma_{Z_{cs}(4000)}=131\pm15\pm26\,\text{MeV}\,,\notag\\
&M_{Z_{cs}(4220)}=4216\pm24^{+43}_{-30}\,\text{MeV}\,,\notag\\
&\Gamma_{Z_{cs}(4220)}=233\pm52^{+97}_{-73}\,\text{MeV}\,,\notag
\end{align}
respectively. Although the mass of $Z_{cs}(4000)$ is very close to that of $Z_{cs}(3985)$,  there is no evidence that $Z_{cs}(4000)$ and $Z_{cs}(3985)$ are the same state~\cite{LHCb 2021}. In present model, the lowest ground $S$-wave $cn\bar{c}\bar{s}$ state with quantum numbers $I(J^{P})=1/2(1^+)$ falls at $\sim3989\,\text{MeV}$, which thus could be interpreted as component of $Z_{cs}(4000)$. Yet, in the mass range $4.16\sim4.26\,\text{GeV}$ for $I(J^{P})=1/2(1^+)$, there are three states whose masses are consistent with the experimental mass of $Z_{cs}(4220)$.

Accordingly, the experimental data of $Z_{cs}$ states, and presently obtained results those are close to the experimental data, are collected in Table~\ref{ZcsCT}.

\begin{table}
\caption{Comparison of model results with experimental masses for $Z_{cs}$ states. Fourth, fifth and sixth columns collect the experimental data of $Z_{cs}$ states, and first two columns collect the model results in which model masses are close to the experimental masses.}\label{ZcsCT}
\renewcommand
\tabcolsep{0.15cm}
\renewcommand{\arraystretch}{1.5}
\begin{tabular}{ccccc}
\hline\hline
\multicolumn{2}{c}{Model values}&\multicolumn{3}{c}{Experiment values}\\
$I(J^P)  $&M\,(MeV) &States                         &$I(J^P)$                   &M\,(MeV)\\
\hline
$1/2(0^+)$&$3951.72$&\multirow{2}{*}{$Z_{cs}(3985)$}&\multirow{2}{*}{$1/2(?^?)$}&\multirow{2}{*}{$3982.5^{+1.8}_{-2.6}\pm2.1$}\\
$1/2(1^+)$&$3988.72$& & &\\
\hline
$1/2(1^+)$&$3988.72$&$Z_{cs}(4000)                 $&$1/2(1^+)                 $&$4003\pm6^{+4}_{-14}$\\
\hline
$1/2(1^+)$&$4165.97$&\multirow{3}{*}{$Z_{cs}(4220)$}&\multirow{3}{*}{$1/2(1^?)$}&\multirow{3}{*}{$4216\pm24^{+43}_{-30}$}\\
$1/2(1^+)$&$4212.73$&  &  &\\
$1/2(1^+)$&$4253.47$&  &  &\\
\hline\hline
\end{tabular}
\end{table}

\subsection{Double-charm $T_{cc}$ states}

Very recently, the first double-charm exotic state $T_{cc}^+$ with $I(J^{P})=0(1^+)$, whose minimal quark content is $cc\bar{u}\bar{d}$, was observed by LHCb~\cite{LHCbTcc1}. Its mass is very close to and below $D^0D^{*+}$ threshold, and its decay width is extremely narrow. A further analysis have been done by LHCb collaboration in Ref.~\cite{LHCbTcc2}.
Theoretically, the tetraquark states containing two heavy quarks have been predicted in Refs.~\cite{Zhu 2013,Kar.M 2017}.

\begin{figure}[htbp]
	\centering
	\includegraphics[scale=0.4]{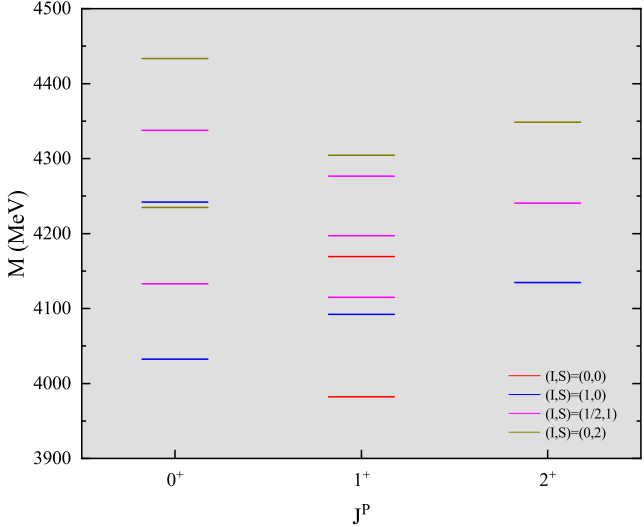}
	\caption{The model spectrum of ground $S$-wave double-charm tetraquark states. Red solid lines denote $T_{cc}^{0,0}$ states, blue solid lines denote $T_{cc}^{1,0}$ states, pink solid lines denote $T_{cc}^{1/2,1}$ states, green solid line denote $T_{cc}^{0,2}$ states.}\label{TccF}
\end{figure}

In present work, the mass spectra of ground $S$-wave $cc\bar{n}\bar{n}$, $cc\bar{n}\bar{s}$ and $cc\bar{s}\bar{s}$ tetraquark systems are studied. The numerical results are listed in Table~\ref{TccT}, where the first three columns are the labels of these states in the model, quark content and quantum number of each state respectively, the fourth, fifth columns show the results of single configuration calculations, and the last two columns give the results obtained by considering configurations mixing effects. The model spectrum of $S$-wave double-charm tetraquark states are also shown in Fig.~\ref{TccF}, where red solid lines denote $T_{cc}^{0,0}$ states, blue solid lines denote $T_{cc}^{1,0}$ states, pink solid lines denote $T_{cc}^{1/2,1}$ states, green solid line denote $T_{cc}^{0,2}$ states.

\begin{table*}[htbp]
\caption{The model results of ground $S$-wave double-charm tetraquark states. First three columns are the labels of these states in the model, quark content and quantum number of each state repectively, the fourth, fifth columns give the results of single configuration calculations, and the sixth, seventh columns give the results after considering configuration mixing.}\label{TccT}
\renewcommand
\tabcolsep{0.45cm}
\renewcommand{\arraystretch}{1.5}
\begin{tabular}{ccccccc}
\hline\hline

  & & &\multicolumn{2}{c}{Single configuration}&\multicolumn{2}{c}{Configurations mixing}\\
 \hline

States                           &Quark content                      &$(I,S)(J^P)                 $&Config.    &M\,(MeV) &M\,(MeV) &Mixing coefficients\\
\hline
\multirow{2}{*}{$T_{cc}^{0,0}$}&\multirow{2}{*}{$cc\bar{n}\bar{n}$}&\multirow{2}{*}{$(0,0)(1^+)$}&$|3\rangle$&$3998.90$&$3982.12$&$(-0.95,\hspace{0.15cm}-0.30)$\\

                                 &                                   &                             &$|6\rangle$&$4152.40$&$4169.18$&$(-0.30,\hspace{0.15cm}0.95)$\\
                                 \hline
\multirow{2}{*}{$T_{cc}^{1,0}$}&\multirow{2}{*}{$cc\bar{n}\bar{n}$}&\multirow{2}{*}{$(1,0)(0^+)$}&$|1\rangle$&$4204.41$&$4032.30$&$(0.42,\hspace{0.15cm}-0.91)$\\

                                 &                                   &                             &$|7\rangle$&$4069.69$&$4241.80$&$(-0.91,\hspace{0.15cm}-0.42)$\\
\hline
$T_{cc}^{1,0}                 $&$cc\bar{n}\bar{n}                 $&$(1,0)(1^+)                 $&$|7\rangle$&$4092.34$&---&---\\
\hline
$T_{cc}^{1,0}                 $&$cc\bar{n}\bar{n}                 $&$(1,0)(2^+)                 $&$|7\rangle$&$4134.59$&---&---\\
\hline
\multirow{2}{*}{$T_{cc}^{\frac12,1}$}&\multirow{2}{*}{$cc\bar{n}\bar{s}$}&\multirow{2}{*}{$(\frac12,1)(0^+)$}&$|1\rangle$&$4296.64$&$4132.95$&$(-0.45,\hspace{0.15cm}-0.89)$\\

                                       &                                   &                                   &$|7\rangle$&$4174.01$&$4337.71$&$(-0.89,\hspace{0.15cm}0.45)$\\
\hline
\multirow{3}{*}{$T_{cc}^{\frac12,1}$}&\multirow{3}{*}{$cc\bar{n}\bar{s}$}&\multirow{3}{*}{$(\frac12,1)(1^+)$}&$|3\rangle$&$4134.14$&$4115.10$&$(-0.94,\hspace{0.15cm}-0.34,\hspace{0.15cm}0.00)$\\

                                       &                                   &                                   &$|6\rangle$&$4257.37$&$4197.18$&$(0.00,\hspace{0.15cm}0.00,\hspace{0.15cm}1.00)$\\

                                       &                                   &                                   &$|7\rangle$&$4197.18$&$4276.41$&$(0.34,\hspace{0.15cm}-0.94,\hspace{0.15cm}0.00)$\\
\hline
$T_{cc}^{\frac12,1}                 $& $cc\bar{n}\bar{s}                $&$(\frac12,1)(2^+)                 $&$|7\rangle$&$4240.55$& ---&---\\
\hline
\multirow{2}{*}{$T_{cc}^{0,2}$}&\multirow{2}{*}{$cc\bar{s}\bar{s}$}&\multirow{2}{*}{$(0,2)(0^+)$}&$|1\rangle$&$4387.45$&$4234.84$&$(0.48,\hspace{0.15cm}-0.88)$\\

                                 &                                   &                             &$|7\rangle$&$4280.69$&$4433.30$&$(-0.88,\hspace{0.15cm}-0.48)$\\
\hline
$T_{cc}^{0,2}                 $&$cc\bar{s}\bar{s}                 $&$(0,2)(1^+)                 $&$|7\rangle$&$4304.25$&---&---\\
\hline
$T_{cc}^{0,2}                 $&$cc\bar{s}\bar{s}                 $&$(0,2)(2^+)                 $&$|7\rangle$&$4348.45$&---&---\\
\hline\hline
\end{tabular}
\end{table*}

As shown in Table~\ref{TccT}, all the presently obtained lowest compact $T_{cc}$ states are higher than thresholds of corresponding charmed meson-meson channels. Therefore, there may be no stable bound states for $S$-wave double-charm tetraquark state. For instance, the lowest $S$-wave $cc\bar{n}\bar{n}$ states with quantum number $I(J^P)=0(1^+)$ in present model falls at $\sim3982\,\text{MeV}$, which is about $100\,\text{MeV}$ higher than the experimental data for $T_{cc}^+$. Thus, $T_{cc}^+$ cannot be interpreted as a compact tetraquark state in our model.

In Ref.~\cite{D.Ebert 2007}, the spectrum of tetraquark states containing double heavy quarks were investigated within a diquark-antidiquark picture employing the relativistic constituent quark model, the obtained numerical results are in agreements with the present ones. Very recently, double-heavy tetraquark states have been also studied using a Chiral quark model in Ref.~\cite{Deng 2021}, it's shown that both the meson exchange force and the coupled channel effects play pivotal roles for reproduction of the binding energy of $T_{cc}^+$ correlative to $D^0D^{*+}$ threshold, revealing the molecular nature of $T_{cc}^+$~\cite{Deng 2021}.

\section{Summary} \label{Summary}

In present work, we study the ground $S$-wave hidden- and double-charm tetraquark states employing a nonrelativistic quark potential model, where the instanton-induced interaction is employed as the spin-dependent residual interaction between quarks. The model parameters are fixed by fitting the spectrum of the ground state hadrons. Our numerical results show that energies for several presently obtained compact tetraquark states are very close to the masses of the experimentally observed meson exotic states.

Particularly, two $c\bar{c}n\bar{n}$ and $c\bar{c}s\bar{s}$ tetraquark states with quantum numbers $I^G(J^{PC})=0^+(1^{++})$
fall at the energies close to $X(4140)$ and $X(4274)$, respectively. Energies of several $c\bar{c}n\bar{n}$ tetraquark states with $I^G(J^{PC})=1^+(1^{+-})$ are close to $Z_c(3900)$, $Z_c(4020)$ and $X(4055)$, and another obtained $c\bar{c}n\bar{n}$ sate with
$I^G(J^{PC})=1^-(0^{++})$ is close to the $X(4100)$ state. In addition, the $X(4050)$ and $X(4250)$ states,
whose quantum numbers have not been determined, may be interpreted as compact $c\bar{c}n\bar{n}$ tetraquark states with quantum numbers $I^G(J^{PC})=1^-(1^{++})$ or $I^G(J^{PC})=1^-(2^{++})$, and $I^G(J^{PC})=1^-(0^{++})$ or $I^G(J^{PC})=1^-(2^{++})$, respectively.

For the exotic meson states with one strange quark, we obtain two states very close to the experimentally observed $Z_{cs}(3985)$ state, which indicates the quantum number of $Z_{cs}(3985)$ may be $J^P=0^+$ or $J^P=1^+$. Furthermore, several $c\bar{c}n\bar{s}$ tetraquark states with $J^P=1^+$ are obtained to fall in the energy range close to the $Z_{cs}(4000)$ and $Z_{cs}(4220)$ states. Accordingly, one may conclude that the presently considered compact tetraquark states my take significant probabilities in the above referred $X$, $Z_c$ and $Z_{cs}$ states.

\begin{acknowledgments}

We would like to thank Prof. X. H. Zhong for very helpful discussions.
This work is partly supported by the Chongqing Natural Science
Foundation under Project No. cstc2021jcyj-msxmX0078,
No. cstc2019jcyj-msxmX0409, and the National
Natural Science Foundation of China under Grant Nos. 12075288, 12075133, 11735003, 11961141012 and 11835015. It is also
supported by the Youth Innovation Promotion Association CAS, Taishan
Scholar Project of Shandong Province (Grant No.tsqn202103062),
the Higher Educational Youth Innovation Science and Technology
Program Shandong Province (Grant No. 2020KJJ004).

\end{acknowledgments}


\begin{thebibliography}{99}	


\bibitem{PDG}
P.~A.~Zyla \textit{et al.} [Particle Data Group],
``Review of Particle Physics,''
PTEP \textbf{2020}, 083C01 (2020).

\bibitem{Belle 2003}
S.~K.~Choi \textit{et al.} [Belle],
``Observation of a narrow charmonium - like state in exclusive $B^{\pm} \to K^{\pm} \pi^+ \pi^- J/\psi$ decays,''
Phys. Rev. Lett. \textbf{91}, 262001 (2003).

\bibitem{BESIII 2013}
M.~Ablikim \textit{et al.} [BESIII],
``Observation of a Charged Charmoniumlike Structure in $e^+e^-$ \textrightarrow{} $\pi^+\pi^-$ J/\ensuremath{\psi} at $\sqrt{s}$ =4.26  GeV,''
Phys. Rev. Lett. \textbf{110}, 252001 (2013).

\bibitem{BESIII 2021}
M.~Ablikim \textit{et al.} [BESIII],
``Observation of a Near-Threshold Structure in the $K^+$ Recoil-Mass Spectra in $e^+e^- \rightarrow K^+(D_s^-D^{*0}+D_s^{*-}D^0$),''
Phys. Rev. Lett. \textbf{126}, 102001 (2021).


\bibitem{Yang:2009zzp}
Y.~Yang, C.~Deng, J.~Ping and T.~Goldman,
``S-wave Q Q anti-q anti-q state in the constituent quark model,''
Phys. Rev. D \textbf{80}, 114023 (2009).

\bibitem{Deng:2016rus}
C.~Deng, J.~Ping, H.~Huang and F.~Wang,
``Heavy pentaquark states and a novel color structure,''
Phys. Rev. D \textbf{95}, 014031 (2017).

\bibitem{Richard:2017una}
J.~M.~Richard, A.~Valcarce and J.~Vijande,
``Stable heavy pentaquarks in constituent models,''
Phys. Lett. B \textbf{774}, 710-714 (2017).

\bibitem{Luo:2017eub}
S.~Q.~Luo, K.~Chen, X.~Liu, Y.~R.~Liu and S.~L.~Zhu,
``Exotic tetraquark states with the $qq\bar{Q}\bar{Q}$ configuration,''
Eur. Phys. J. C \textbf{77}, 709 (2017).

\bibitem{JinX 2021}
X.~Jin, Y.~Wu, X.~Liu, Y.~Xue, H.~Huang, J.~Ping and B.~Zhong,
``Strange hidden-charm tetraquarks in constituent quark model,''
Eur. Phys. J. C \textbf{81}, 1108 (2021).

\bibitem{YangG 2021}
G.~Yang, J.~Ping and J.~Segovia,
``Hidden-charm tetraquarks with strangeness in the chiral quark model,''
Phys. Rev. D \textbf{104}, 094035 (2021).

\bibitem{Deng 2021}
C.~Deng and S.~L.~Zhu,
``$T^+_{cc}$ and its partners,''
[arXiv:2112.12472 [hep-ph]].



\bibitem{Chen:2013pya}
W.~Chen, H.~y.~Jin, R.~T.~Kleiv, T.~G.~Steele, M.~Wang and Q.~Xu,
``QCD sum-rule interpretation of X(3872) with $J^{PC}=1^{++}$ mixtures of hybrid charmonium and $\bar{D}D^*$ molecular currents,''.
Phys. Rev. D \textbf{88}, 045027 (2013)

\bibitem{Chen:2015ata}
W.~Chen, T.~G.~Steele, H.~X.~Chen and S.~L.~Zhu,
``Mass spectra of $Z_c$ and $Z_b$ exotic states as hadron molecules,''
Phys. Rev. D \textbf{92}, 054002 (2015).

\bibitem{Wang:2017dtg}
Z.~G.~Wang and Z.~H.~Yan,
``Analysis of the scalar, axialvector, vector, tensor doubly charmed tetraquark states with QCD sum rules,''
Eur. Phys. J. C \textbf{78}, 19 (2018).

\bibitem{Wang:2019got}
Z.~G.~Wang,
``Analysis of the $P_c(4312)$, $P_c(4440)$, $P_c(4457)$ and related hidden-charm pentaquark states with QCD sum rules,''
Int. J. Mod. Phys. A \textbf{35}, 2050003 (2020).


\bibitem{D.Ebert 2007}
D.~Ebert, R.~N.~Faustov, V.~O.~Galkin and W.~Lucha,
``Masses of tetraquarks with two heavy quarks in the relativistic quark model,''
Phys. Rev. D \textbf{76}, 114015 (2007).

\bibitem{Esposito:2013fma}
A.~Esposito, M.~Papinutto, A.~Pilloni, A.~D.~Polosa and N.~Tantalo,
``Doubly charmed tetraquarks in $B_c$ and $\Xi_{bc}$ decays,''
Phys. Rev. D \textbf{88}, no.5, 054029 (2013)

\bibitem{Shi:2021jyr}
P.~P.~Shi, F.~Huang and W.~L.~Wang,
``Hidden charm tetraquark states in a diquark model,''
Phys. Rev. D \textbf{103}, 094038 (2021).


\bibitem{Liu:2009qhy}
X.~Liu, Z.~G.~Luo, Y.~R.~Liu and S.~L.~Zhu,
``X(3872) and Other Possible Heavy Molecular States,''
Eur. Phys. J. C \textbf{61}, 411-428 (2009).

\bibitem{WangP 2013}
P.~Wang and X.~G.~Wang,
``Study on $X(3872)$ from effective field theory with pion exchange interaction,''
Phys. Rev. Lett. \textbf{111}, 042002 (2013).

\bibitem{HeJ 2013}
J.~He, X.~Liu, Z.~F.~Sun and S.~L.~Zhu,
``$Z_c(4025)$ as the hadronic molecule with hidden charm,''
Eur. Phys. J. C \textbf{73}, 2635 (2013).

\bibitem{Aceti:2014kja}
F.~Aceti, M.~Bayar, J.~M.~Dias and E.~Oset,
``Prediction of a $Z_c(4000)$ $D^* \bar D^*$ state and relationship to the claimed $Z_c(4025)$,''
Eur. Phys. J. A \textbf{50}, 103 (2014).

\bibitem{ChenR 2021}
R.~Chen and Q.~Huang,
``$Z_{cs}(3985)^-$: A strange hidden-charm tetraquark resonance or not?,''
Phys. Rev. D \textbf{103}, 034008 (2021).

\bibitem{Feijoo:2021ppq}
A.~Feijoo, W.~H.~Liang and E.~Oset,
``D0D0\ensuremath{\pi}+ mass distribution in the production of the Tcc exotic state,''
Phys. Rev. D \textbf{104}, no.11, 114015 (2021)


\bibitem{Chen:2016qju}
H.~X.~Chen, W.~Chen, X.~Liu and S.~L.~Zhu,
``The hidden-charm pentaquark and tetraquark states,''
Phys. Rept. \textbf{639}, 1-121 (2016).

\bibitem{Lebed:2016hpi}
R.~F.~Lebed, R.~E.~Mitchell and E.~S.~Swanson,
``Heavy-Quark QCD Exotica,''
Prog. Part. Nucl. Phys. \textbf{93}, 143-194 (2017).

\bibitem{Guo:2017jvc}
F.~K.~Guo, C.~Hanhart, U.~G.~Mei\ss{}ner, Q.~Wang, Q.~Zhao and B.~S.~Zou,
``Hadronic molecules,''
Rev. Mod. Phys. \textbf{90}, 015004 (2018).

\bibitem{Liu:2019zoy}
Y.~R.~Liu, H.~X.~Chen, W.~Chen, X.~Liu and S.~L.~Zhu,
``Pentaquark and Tetraquark states,''
Prog. Part. Nucl. Phys. \textbf{107}, 237-320 (2019).

\bibitem{Brambilla:2019esw}
N.~Brambilla, S.~Eidelman, C.~Hanhart, A.~Nefediev, C.~P.~Shen, C.~E.~Thomas, A.~Vairo and C.~Z.~Yuan,
``The $XYZ$ states: experimental and theoretical status and perspectives,''
Phys. Rept. \textbf{873}, 1-154 (2020).


\bibitem{LHCbTcc1}
R.~Aaij \textit{et al.} [LHCb],
``Observation of an exotic narrow doubly charmed tetraquark,''
[arXiv:2109.01038 [hep-ex]].

\bibitem{LHCbTcc2}
R.~Aaij \textit{et al.} [LHCb],
``Study of the doubly charmed tetraquark $T_{cc}^+$,''
[arXiv:2109.01056 [hep-ex]].

\bibitem{Zhu 2013}
N.~Li, Z.~F.~Sun, X.~Liu and S.~L.~Zhu,
``Coupled-channel analysis of the possible $D^{(*)}D^{(*)}, \bar{B}^{(*)}\bar{B}^{(*)}$ and $D^{(*)}\bar{B}^{(*)}$ molecular states,''
Phys. Rev. D \textbf{88}, 114008 (2013).

\bibitem{Kar.M 2017}
M.~Karliner and J.~L.~Rosner,
``Discovery of doubly-charmed $\Xi_{cc}$ baryon implies a stable ($b b \bar{u} \bar{d}$) tetraquark,''
Phys. Rev. Lett. \textbf{119}, 202001 (2017).

\bibitem{Xicc}
R.~Aaij \textit{et al.} [LHCb],
``Observation of the doubly charmed baryon $\Xi_{cc}^{++}$,''
Phys. Rev. Lett. \textbf{119}, 112001 (2017).

\bibitem{Ling:2021bir}
X.~Z.~Ling, M.~Z.~Liu, L.~S.~Geng, E.~Wang and J.~J.~Xie,
``Can we understand the decay width of the $T_{cc}^+$ state?,''
Phys. Lett. B \textbf{826}, 136897 (2022).

%
\bibitem{Isgur}
S.~Godfrey and N.~Isgur,
``Mesons in a Relativized Quark Model with Chromodynamics,''
Phys. Rev. D \textbf{32}, 189-231 (1985).

\bibitem{Glozman:1995fu}
L.~Y.~Glozman and D.~O.~Riska,
``The Spectrum of the nucleons and the strange hyperons and chiral dynamics,''
Phys. Rept. \textbf{268}, 263-303 (1996).

\bibitem{Glozman:1995xy}
L.~Y.~Glozman and D.~O.~Riska,
``The Charm and bottom hyperons and chiral dynamics,''
Nucl. Phys. A \textbf{603}, 326-344 (1996)
[erratum: Nucl. Phys. A \textbf{620}, 510-510 (1997)].

\bibitem{Vijande:2004he}
J.~Vijande, F.~Fernandez and A.~Valcarce,
``Constituent quark model study of the meson spectra,''
J. Phys. G \textbf{31}, 481 (2005).

\bibitem{Deng:2014gqa}
C.~Deng, J.~Ping and F.~Wang,
``Interpreting $Z_c(3900)$ and $Z_c(4025)/Z_c(4020)$ as charged tetraquark states,''
Phys. Rev. D \textbf{90}, 054009 (2014).

\bibitem{Yang:2015bmv}
G.~Yang and J.~Ping,
``The structure of pentaquarks $P_c^+$ in the chiral quark model,''
Phys. Rev. D \textbf{95}, 014010 (2017).

\bibitem{Deng:2017xlb}
C.~Deng, J.~Ping, H.~Huang and F.~Wang,
``Hidden charmed states and multibody color flux-tube dynamics,''
Phys. Rev. D \textbf{98}, 014026 (2018).

\bibitem{Yuan:2012wz}
S.~G.~Yuan, K.~W.~Wei, J.~He, H.~S.~Xu and B.~S.~Zou,
``Study of $qqqc\bar{c}$ five quark system with three kinds of quark-quark hyperfine interaction,''
Eur. Phys. J. A \textbf{48}, 61 (2012).

\bibitem{Yuan:2012zs}
S.~G.~Yuan, C.~S.~An, K.~W.~Wei, B.~S.~Zou and H.~S.~Xu,
``Spectrum of low-lying $s^{3}Q\bar{Q}$ configurations with negative parity,''
Phys. Rev. C \textbf{87}, 025205 (2013).

\bibitem{An:2013zoa}
C.~S.~An, B.~C.~Metsch and B.~S.~Zou,
``Mixing of the low-lying three- and five-quark $\Omega$ states with negative parity,''
Phys. Rev. C \textbf{87}, 065207 (2013).

\bibitem{An:2014lga}
C.~S.~An and B.~S.~Zou,
``Low-lying $\Omega$ states with negative parity in an extended quark model with Nambu-Jona-Lasinio interaction,''
Phys. Rev. C \textbf{89}, 055209 (2014).

\bibitem{An:2017lwg}
C.~S.~An and H.~Chen,
``Observed $\Omega_{c}^{0}$ resonances as pentaquark states,''
Phys. Rev. D \textbf{96}, 034012 (2017).

\bibitem{Wang:2021rjk}
J.~B.~Wang, G.~Li, C.~R.~Deng, C.~S.~An and J.~J.~Xie,
``\ensuremath{\Omega}cc resonances with negative parity in the chiral constituent quark model,''
Phys. Rev. D \textbf{104}, 094008 (2021).


\bibitem{tHooft}
G.~'t Hooft,
``Computation of the Quantum Effects Due to a Four-Dimensional Pseudoparticle,''
Phys. Rev. D \textbf{14}, 3432-3450 (1976)
[erratum: Phys. Rev. D \textbf{18}, 2199 (1978)].

\bibitem{Diakonov:1983hh}
D.~Diakonov and V.~Y.~Petrov,
``Instanton Based Vacuum from Feynman Variational Principle,''
Nucl. Phys. B \textbf{245}, 259-292 (1984).

\bibitem{Diakonov:1985eg}
D.~Diakonov and V.~Y.~Petrov,
``A Theory of Light Quarks in the Instanton Vacuum,''
Nucl. Phys. B \textbf{272}, 457-489 (1986).

\bibitem{Shuryak:1988bf}
E.~V.~Shuryak and J.~L.~Rosner,
``Instantons and Baryon Mass Splittings,''
Phys. Lett. B \textbf{218}, 72-74 (1989).

\bibitem{Blask:1990ez}
W.~H.~Blask, U.~Bohn, M.~G.~Huber, B.~C.~Metsch and H.~R.~Petry,
``Hadron spectroscopy with instanton induced quark forces,''
Z. Phys. A \textbf{337}, 327-335 (1990).

\bibitem{Brau:1998sxe}
F.~Brau and C.~Semay,
``Light meson spectra and instanton induced forces,''
Phys. Rev. D \textbf{58}, 034015 (1998).

\bibitem{Semay:2001th}
C.~Semay, F.~Brau and B.~Silvestre-Brac,
``Baryon spectra with instanton induced forces,''
Phys. Rev. C \textbf{64}, 055202 (2001).

\bibitem{Migura:2006ep}
S.~Migura, D.~Merten, B.~Metsch and H.~R.~Petry,
``Charmed baryons in a relativistic quark model,''
Eur. Phys. J. A \textbf{28}, 41 (2006).

\bibitem{Metch}
M.~W.~Beinker, B.~C.~Metsch and H.~R.~Petry,
``Bound $q^2$-anti-$q^2$ states in a constituent quark model,''
J. Phys. G \textbf{22}, 1151-1160 (1996).

\bibitem{tHooft:2008rus}
G.~'t Hooft, G.~Isidori, L.~Maiani, A.~D.~Polosa and V.~Riquer,
``A Theory of Scalar Mesons,''
Phys. Lett. B \textbf{662}, 424-430 (2008)


\bibitem{Cornell}
E.~Eichten, K.~Gottfried, T.~Kinoshita, J.~B.~Kogut, K.~D.~Lane and T.~M.~Yan,
``The Spectrum of Charmonium,''
Phys. Rev. Lett. \textbf{34}, 369-372 (1975)
[erratum: Phys. Rev. Lett. \textbf{36}, 1276 (1976)].




\bibitem{Zhang:2007mu}
M.~Zhang, H.~X.~Zhang and Z.~Y.~Zhang,
``$QQ$ anti-$q$ anti-$q$ four-quark bound states in chiral SU(3) quark model,''
Commun. Theor. Phys. \textbf{50}, 437-440 (2008).

\bibitem{Zhang:2005jz}
D.~Zhang, F.~Huang, Z.~Y.~Zhang and Y.~W.~Yu,
``Further study on $5q$ configuration states in the chiral SU(3) quark model,''
Nucl. Phys. A \textbf{756}, 215-226 (2005).

\bibitem{Liu:2019zuc}
M.~S.~Liu, Q.~F.~L\"u, X.~H.~Zhong and Q.~Zhao,
``All-heavy tetraquarks,''
Phys. Rev. D \textbf{100}, 016006 (2019).

\bibitem{GEM}
E.~Hiyama, Y.~Kino and M.~Kamimura,
``Gaussian expansion method for few-body systems,''
Prog. Part. Nucl. Phys. \textbf{51}, 223-307 (2003).

\bibitem{tHooft:1999cta}
G.~'t Hooft,
``The Physics of instantons in the pseudoscalar and vector meson mixing,''
[arXiv:hep-th/9903189 [hep-th]].

\bibitem{LHCb 2013}
R.~Aaij \textit{et al.} [LHCb],
``Determination of the $X(3872)$ meson quantum numbers,''
Phys. Rev. Lett. \textbf{110}, 222001 (2013).

\bibitem{Belle 2006}
S.~Uehara \textit{et al.} [Belle],
``Observation of a $\chi'_{c2}$ candidate in $\gamma \gamma \to D \bar{D}$ production at BELLE,''
Phys. Rev. Lett. \textbf{96}, 082003 (2006).

\bibitem{Belle 2010}
S.~Uehara \textit{et al.} [Belle],
``Observation of a charmonium-like enhancement in the $\gamma \gamma \to \omega J/\psi$ process,''
Phys. Rev. Lett. \textbf{104}, 092001 (2010).

\bibitem{BaBar 2012}
J.~P.~Lees \textit{et al.} [BaBar],
``Study of $X(3915) \to J/\psi \omega$ in two-photon collisions,''
Phys. Rev. D \textbf{86}, 072002 (2012).

\bibitem{CDF 2009}
T.~Aaltonen \textit{et al.} [CDF],
``Evidence for a Narrow Near-Threshold Structure in the $J/\psi\phi$ Mass Spectrum in $B^+\to J/\psi\phi K^+$ Decays,''
Phys. Rev. Lett. \textbf{102}, 242002 (2009).

\bibitem{CDF 2011}
T.~Aaltonen \textit{et al.} [CDF],
``Observation of the $Y(4140)$ Structure in the $J/\psi\phi$ Mass Spectrum in $B^\pm\to J/\psi\phi K^\pm$ Decays,''
Mod. Phys. Lett. A \textbf{32}, 1750139 (2017).

\bibitem{LHCb 2017}
R.~Aaij \textit{et al.} [LHCb],
``Observation of $J/\psi\phi$ structures consistent with exotic states from amplitude analysis of $B^+\to J/\psi \phi K^+$ decays,''
Phys. Rev. Lett. \textbf{118}, 022003 (2017).

\bibitem{LHCb PRD 2017}
R.~Aaij \textit{et al.} [LHCb],
``Amplitude analysis of $B^+\to J/\psi \phi K^+$ decays,''
Phys. Rev. D \textbf{95}, 012002 (2017).

\bibitem{LHCb 2021}
R.~Aaij \textit{et al.} [LHCb],
``Observation of New Resonances Decaying to $J/\psi K^+$ and $J/\psi \phi$,''
Phys. Rev. Lett. \textbf{127}, 082001 (2021).

\bibitem{Belle 2013}
Z.~Q.~Liu \textit{et al.} [Belle],
``Study of $e^+e^-\rightarrow\pi^+\pi^-J/\Psi$ and Observation of a Charged Charmoniumlike State at Belle,''
Phys. Rev. Lett. \textbf{110}, 252002 (2013)
[erratum: Phys. Rev. Lett. \textbf{111}, 019901 (2013)].

\bibitem{BESIII 2014}
M.~Ablikim \textit{et al.} [BESIII],
``Observation of a charged $(D\bar{D}^{*})^\pm$ mass peak in $e^{+}e^{-} \to \pi D\bar{D}^{*}$ at $\sqrt{s} =$ 4.26 GeV,''
Phys. Rev. Lett. \textbf{112}, 022001 (2014).

\bibitem{BESIII 2014 Zc4025}
M.~Ablikim \textit{et al.} [BESIII],
``Observation of a charged charmoniumlike structure in $e^+e^- \to (D^{*} \bar{D}^{*})^{\pm} \pi^\mp$ at $\sqrt{s}=4.26$GeV,''
Phys. Rev. Lett. \textbf{112}, 132001 (2014).


\bibitem{BESIII 2013 Zc4020}
M.~Ablikim \textit{et al.} [BESIII],
``Observation of a Charged Charmoniumlike Structure $Z_c$(4020) and Search for the $Z_c$(3900) in $e^+e^-\rightarrow\pi^+\pi^-h_c$,''
Phys. Rev. Lett. \textbf{111}, 242001 (2013).

\bibitem{Belle 2008}
R.~Mizuk \textit{et al.} [Belle],
``Observation of two resonance-like structures in the $\pi^+ \chi_{c1}$ mass distribution in exclusive $\bar{B}^0 \to K^- \pi^+ \chi_{c1}$ decays,''
Phys. Rev. D \textbf{78}, 072004 (2008).

\bibitem{Belle 2015}
X.~L.~Wang \textit{et al.} [Belle],
``Measurement of $e^+e^- \to \pi^+\pi^-\psi(2S)$ via Initial State Radiation at Belle,''
Phys. Rev. D \textbf{91}, 112007 (2015).

\bibitem{LHCb 2018}
R.~Aaij \textit{et al.} [LHCb],
``Evidence for an $\eta _c(1S) \pi ^-$ resonance in $B^0 \rightarrow \eta _c(1S) K^+\pi ^-$ decays,''
Eur. Phys. J. C \textbf{78}, 1019 (2018).

\bibitem{Guo 2021}
Z.~H.~Guo and J.~A.~Oller,
``Unified description of the hidden-charm tetraquark states $Z_{cs}(3985)$, $Z_c(3900)$, and $X(4020)$,''
Phys. Rev. D \textbf{103}, 054021 (2021).

\bibitem{Ikeno:2021mcb}
N.~Ikeno, R.~Molina and E.~Oset,
``$Z_{cs}$ states from the $D^*_s \overline{D}$ and $J/\Psi K^*$ coupled channels: Signal in $B^+ \rightarrow J/\Psi \tau K^*$ decay,''
Phys. Rev. D \textbf{105}, no.1, 014012 (2022)




\end{thebibliography}
\end{document}